\documentclass[reprint,aps,pra,twocolumn,tightenlines,floatfix,showpacs,superscriptaddress,10pt]{revtex4-1}

\usepackage{amsmath}
\usepackage{amsfonts}
\usepackage{amssymb}
\usepackage{fancyvrb}
\usepackage{graphicx}
\usepackage{color}
\usepackage{titlesec}
\usepackage[normalem]{ulem}

\global\long\def\im{\imath}

\newcommand{\dg} {\dagger}
\newcommand{\pd} {{\phantom\dagger}}

\newcommand{\ci}[1] {c_{#1}^\pd}
\newcommand{\cid}[1] {c_{#1}^\dg}

\newcommand{\qc} {\mathcal{C}}
\newcommand{\qs} {\mathcal{S}}
\newcommand{\qp} {\mathcal{P}}
\newcommand{\qt} {\mathcal{T}}

\renewcommand{\Re}{\operatorname{Re}}
\renewcommand{\Im}{\operatorname{Im}}

\newcommand{\dex} {D_{\text{ex}}}
\newcommand{\TT} {\left| \qt \right|}

\def\bibtexbib{0} 
\graphicspath{{}} 

\begin{document}

\author{Daniel Gruss}

\affiliation{Center for Nanoscale Science and Technology,
             National Institute of Standards and Technology,
             Gaithersburg, MD 20899}
\affiliation{Maryland NanoCenter, University of Maryland,
             College Park, MD 20742}
\affiliation{Department of Physics, Oregon State University,
             Corvallis, OR 97331}

\author{Chih-Chun Chien} 

\affiliation{School of Natural Sciences, University of California,
             Merced, CA 95343}
            
\author{Julio T. Barreiro}

\affiliation{Department of Physics, University of California,
	San Diego, CA, 92093}

\author{Massimiliano Di Ventra}

\affiliation{Department of Physics, University of California,
             San Diego, CA, 92093}

\author{Michael Zwolak} \email[Contact: ]{mpz@nist.gov}

\affiliation{Center for Nanoscale Science and Technology,
             National Institute of Standards and Technology,
             Gaithersburg, MD 20899}
             
\affiliation{Biophysics Group, Microsystems \& Nanotechnology Division, 
Physical Measurement Laboratory, National Institute of Standards and Technology, 
Gaithersburg, MD 20899}

\title{An energy-resolved atomic scanning probe}

\begin{abstract}
We propose a method to probe the local density of states (LDOS) of atomic systems that provides both spatial and energy resolution. The method combines atomic and tunneling techniques to supply a simple, yet quantitative and operational, definition of the LDOS for both interacting and non-interacting systems: It is the rate at which particles can be siphoned from the system of interest by a narrow energy band of non-interacting states contacted locally to the many-body system of interest. Ultracold atoms in optical lattices are a natural platform for implementing this broad concept to visualize the energy and spatial dependence of the atom density in interacting, inhomogeneous lattices. This includes models of strongly correlated condensed matter systems, as well as ones with non-trivial topologies.
\end{abstract}

\maketitle

\section{Introduction}

The scanning tunneling microscope (STM)~\cite{binnig1982surface, hansma1987scanning, chen1993introduction, stroscio1993methods, wiesendanger1994scanning} is arguably the most versatile instrument for probing the {\em local} density of states (LDOS) of material surfaces, molecules, and devices. However, there are physical limitations on the information that can be retrieved from such a system due to the scale of the device and the lack of tunable atomic parameters. While measuring the LDOS at the position of the probe tip, the scanning probes measure the LDOS at the position of the probe tip, but do not have access to the whole energy spectrum. Although alternative spectroscopic techniques overcome this limitation, they do not generally provide spatial resolution. Cold-atoms provide many-body, tunable systems that allow for physically simulating models of traditional solid-state materials~\cite{bloch2008many, fertig2005strongly, strohmaier2007interaction, Lamacraft12, chien2015quantum}, including at timescales suitable for quantum transport~\cite{ott2004collisionally, gunter2006bose, chien2012bosonic,brantut2012conduction,chien2013interaction,chien2014landauer, krinner2015observation}. These systems allow for a direct simulation of electronic current through the motion of fermionic atoms in an artificially generated potential. Thus, they give a platform to study -- in a controllable manner -- strongly correlated materials or those with many relevant interactions and treat issues difficult for solid-state experiments and simulation.

The density of states is ubiquitous in classical and quantum physics, as it quantifies the energy distribution of available states. In particular, 
the \emph{local} density of states (LDOS) gives the available states at position $\mathbf{r}$ and frequency $\omega$, 
\begin{equation}
D(\mathbf{r},\omega)=\sum_{n} |\langle \mathbf{r} | \phi_n \rangle|^2 
                              \delta(\omega-\omega_n).
\end{equation}
Here, $|\phi_n\rangle$ is the $n$-th eigenfunction of the full Hamiltonian with energy eigenvalue $\hbar \omega_n$ and $\hbar$ is the reduced Planck's constant. Direct evaluation and probing of the LDOS, however, are challenging in interacting systems~\cite{Hedin70}, so one usually resorts to an {\it operational} definition. In the standard operation of a scanning tunneling microscope (STM), to probe the LDOS $D(\mu)$ at energy $\hbar \mu$ (and implicitly at the position of the tip), the tip distance is held constant while the sample voltage bias $-V$ is changed. This leads to a steady-state current~\cite{hansma1987scanning, chen1993introduction, stroscio1993methods,wiesendanger1994scanning},
\begin{equation}
I \propto \int_{-eV/\hbar}^0 D(\omega) d\omega . \label{eq:STM1}
\end{equation}
The LDOS is then found from the differential conductance,
\begin{equation}
D(\mu) \propto \left. \frac{d I}{dV} \right|_{eV/\hbar=\mu} . \label{eq:STM2}
\end{equation}
This expression neglects the voltage dependence of the electronic transmission from tip to sample, among other factors. Interpretation issues notwithstanding, this limits the accurate extraction of the native electronic density of states to the linear response regime, as not all changes in the current are due to changes in the LDOS. In particular, when the system has strong many-body interactions, such as a poorly screened electronic impurity, a large applied bias will disturb the natural local state by disrupting the nearby electron density. 
 
\begin{figure}[t]
\includegraphics[width=\linewidth]{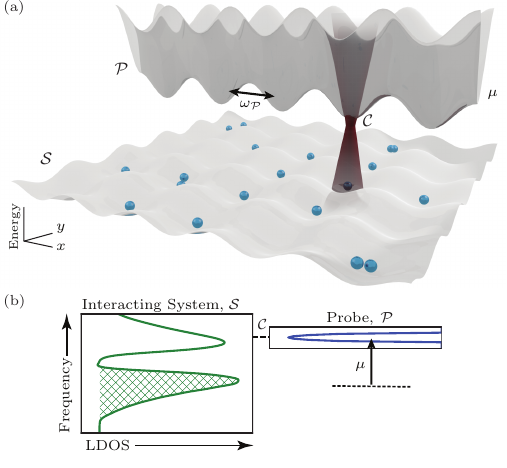}
\caption{The many-body LDOS of an interacting system $\qs$ is measured by putting it ``in contact'' with a non-interacting, narrow band probe $\qp$ via the focused laser beam $\qc$. (a) The cold-atom system scenario can be realized by optical tweezers or optical lattice potentials, as shown schematically, with alkaline-earth atoms~\cite{stellmer2014}. The lattice laser beams' intensities and dimensions determine the coupling rates and trapping frequencies.
The system (probe) lattice traps only atoms in their ground (metastable excited) electronic state~\cite{prl-101-170504}. Coupling between the probe and system lattices is driven by a laser beam (i) tuned to the transition between the two electronic states of the atoms, and (ii) focused on the sites where the atoms ``flow'' between $\qs$ and $\qp$. The probing band has a frequency offset $\mu$ and a small hopping frequency $\omega_\qp$. (b) A representation of the setup in frequency space shows the narrower range of frequencies and offset of the probe. We focus  on one-dimensional systems in this work, though this technique applies to any number of dimensions. \label{fig:expdiagram}}
\end{figure}

Here, we use the tunability of cold-atom lattices to give a direct protocol for LDOS measurement and interpretation. This allows, for instance, the characterization of energy states and the prediction and delineation of contributions to steady-state currents. These systems provide a means to simulate condensed matter~\cite{bloch2008many, fertig2005strongly, strohmaier2007interaction, Lamacraft12, chien2015quantum}, including transport phenomena~\cite{ott2004collisionally, gunter2006bose, brantut2012conduction, krinner2015observation}, while simultaneously yielding opportunities to go beyond solid-state scenarios. In this vein, rather than an applied bias, we propose to use a narrow band probe $\qp$ that scans in energy to interrogate the many-body system $\qs$ (see Fig.~\ref{fig:expdiagram}). The particle current into (out of) an empty (full) reservoir of bandwidth $4\omega_\qp$ offset to a frequency $\mu$ is proportional to the LDOS, i.e., the fraction of particles at frequency $\mu$, 
\begin{equation} \label{eq:propcurr}
I  \approx - 2 \omega_\qp \int_{\mu - 2\omega_\qp}^{\mu + 2\omega_\qp}
                                D(\omega) d\omega 
   \approx -8\omega_\qp^2 D(\mu)\,,
\end{equation}
so long as $\omega_\qp$ is small relative to variations of $D(\omega)$. The occupied (unoccupied) local density of states is given by using an empty (full) probe $\qp$. 

This setup requires tunability of $\qp$: Its chemical potential, occupation, bandwidth/hopping, and ``contact'' magnitude/location with $\qs$ need to be adjustable without compromising its non-interacting behavior. Ultra-cold atoms in artificial lattice potentials are naturally suited for this, see Fig.~\ref{fig:expdiagram}(a) for a schematic. In this context, the method is applicable to a wider class of systems and simulations than spatially resolved radio-frequency spectroscopy, a powerful, yet invasive, tomographic method better suited to homogeneous systems with appropriate atomic hyperfine electronic states and photodissociative mechanisms~\cite{prl-90-230404,prl-99-090403}.

\section{Results}

\subsection{General Approach}

The Hamiltonian, $H$, of the system and probe is
\begin{equation} \label{eq:tothamil}
H_\qs + H_\qc - \hbar \omega_\qp \sum_{i \in \qp} \left( \cid{i} \ci{i+1} + 
                                                         \text{h.c.} \right) 
      + \hbar \mu \sum_{i \in \qp} \cid{i} \ci{i},
\end{equation}
where $\omega_\qp$ and $\mu$ are the hopping and \emph{relative} onsite frequencies, respectively (i.e., the trapping frequencies for a cold-atom system) and $\cid{i}$ ($\ci{i}$) are the creation (annihilation) operators of site $i$. The many-body Hamiltonian of $\qs$ is $H_\qs$ , and the contact Hamiltonian between $\qs$ and $\qp$ is  $H_\qc$, which we also take to have hopping frequency $\omega_\qp$, giving a weak tunneling into a narrow band. Although not necessary, for simplicity, the probe and systems are taken to be one dimensional.

Ultracold atoms in dynamically generated optical lattices and tweezers are ideal for implementing the concept above. Independent manipulation of atoms in their ground and excited (metastable) electronic configurations allows for the construction of the system and probe~\cite{prl-101-170504}. Controllably overlapping the system, probe, and coupling light fields can be readily implemented with feedback-controlled tweezers and lattices generated by deformable micro-mirror devices and spatial light modulators~\cite{s-354-1021, s-354-1024}. After $\qs$ and $\qp$ are loaded with ultracold atoms in different electronic configurations and in their lattice ground states, laser beam $\theta t$-pulses of duration $t$ and phase $\delta$ focused on sites $s$ and $p$ will activate the contact between $\qs$ and $\qp$: $H_\qc=\hbar\omega_\qp' a^\dagger_p a_s + \text{h.c.}$ with $\omega_\qp'=i\theta e^{i\delta}$. In principle, $\omega_\qp'$ can be tuned, but in the following analysis we take $\omega_\qp'=\omega_\qp$ for simplicity. The composite system then evolves, giving rise to the current $I$, which yields the LDOS at frequency $\mu$ according to Eq.~\eqref{eq:propcurr}. The measurement is repeated for each $\mu$ of interest. Furthermore, the system could be probed at multiple contact points, repeatedly, and using several probes
. The detection of the tunneling current can be achieved by monitoring the particle density in the probe~\cite{PhysRevLett.120.103201}, whose time derivative is the particle current exchanged between the system and the probe.

Any experimental realization of this method requires tunability of hopping frequencies (the probe bandwidth), onsite frequencies (the probe offset/chemical potential), among other parameters. The trap depth and spacing controlling the hopping can be tuned via the magnitude of the trapping potential. The onsite frequency is related to the ground state of each well, which can be adjusted with the transverse size of the lattice beam
. The chemical potential is tuned via the external trapping potential, and is also influenced by the particle density. In $\qs$, Bose-Fermi mixtures can be loaded or the optical potential engineered beyond a lattice to allow for the simulation of different types of many-body systems, such as spatially inhomogeneous or varying strength interactions. Further control, including interactions via optical or orbital Feshbach resonances~\cite{prl-115-135301, prl-115-265302, prl-115-265301}, can be achieved by assembling the two non-overlapping lattices atom by atom~\cite{s-354-1021, s-354-1024}. 

\subsection{Non-interacting LDOS}

\begin{figure}
\includegraphics[width=\linewidth]{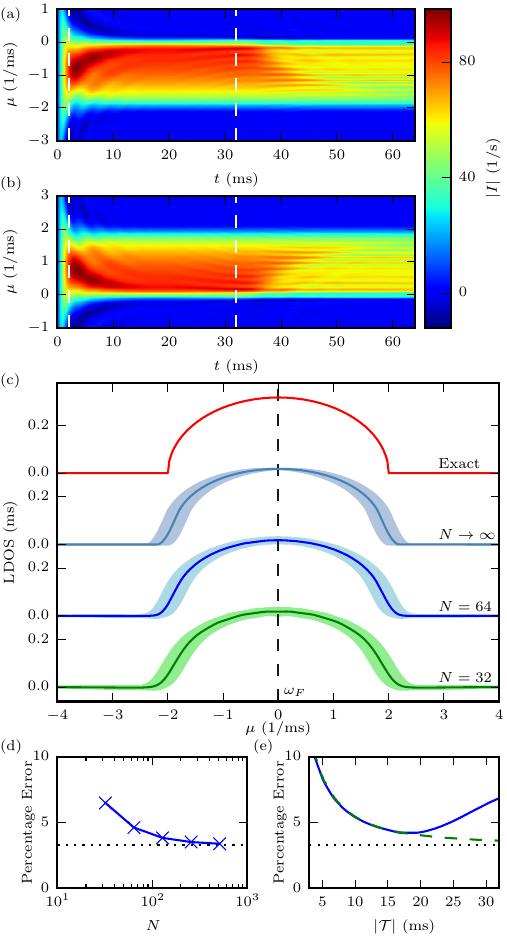}
\caption{The current $I$ versus time $t$ for the (a) occupied and (b) unoccupied $\qs$ states using an initially empty and filled $\qp$, respectively, offset to frequency $\mu$. Here, $\omega_\qs = 1$ ms$^{-1}$ (a typical cold-atom tunneling frequency), $\omega_\qp=0.1$ ms$^{-1}$, and the total lattice length is $N=64$. (c) The LDOS for $\qs$ determined from Eq.~\eqref{eq:propcurr} using the average current in the region $t \in \qt=[2$~ms, $32$~ms$]$ [demarcated by dashed lines in (a) and (b)], which minimizes transient and edge effects. The error bars (shaded regions) indicate the standard deviation of the current in the region $\qt$ combined with the  broadening error 2$\omega_\qp$. The Fermi level, $\omega_F$, is found from the point where the occupied and unoccupied states cross over. (d) Integrated error of the LDOS versus total lattice size and (e) averaging time [for $N=32$ (blue solid line) and $N=512$ (green dashed line)]. The baseline error [when $N\to\infty$, the dotted line in (d) and (e)] is set by the non-zero $\omega_\qp$, which broadens the actual LDOS [see Eq.~\eqref{eq:toterror}]. For short lattices and times, the error in the LDOS is already only a few percent. Thus, even modest-sized systems or calculations can effectively reconstruct the LDOS. \label{fig:avecurrnonint}}
\end{figure}

We illustrate this method by examining a many-body system with Hamiltonian
\begin{equation}
H_\qs = -\hbar \sum_{i \in \qs, \sigma} \omega_{i}
                \left( \cid{i,\sigma} \ci{i+1,\sigma} + \text{h.c.} \right) 
         + \hbar \sum_{i \in \qs} U_i n_{i, \uparrow} n_{i, \downarrow} ,
\end{equation}
where $\omega_{i}$ are tunneling frequencies in the system (sometimes taken to be uniform, i.e., $\omega_i = \omega_\qs$), $n_{i,\sigma}=\cid{i,\sigma} \ci{i,\sigma}$ is the number operator, and $U_i$ is the interaction frequency. The two (or more) components may refer to the spins of electrons or internal states of ultracold atoms. $H_\qp$ can also be expanded in a similar manner to include spin. 

We first examine a spin-polarized, non-interacting system $\qs$: $U_i=0$ and $\omega_{i}=\omega_\qs$ for all $i \in \qs$, which can be both solved exactly for the current \cite{Zwolak04-1,chien2012bosonic,chien2014landauer} and the LDOS (see Appendix~\ref{sec:nonintldos}). Figure~\ref{fig:avecurrnonint}(c) shows the reconstruction of the LDOS using Eq.~\eqref{eq:propcurr} with a finite lattice and time average, giving quantitative agreement with the exact LDOS. The dominant source of error is the finite probe bandwidth $\omega_\qp$, not the length of the lattice or the time of the average [Figs.~\ref{fig:avecurrnonint}(d) and (e)]. This demonstrates that cold-atom systems or many-body simulations are well suited to implement this method, even though they are limited to finite lengths and times. As the total lattice length, $N \to \infty$, the averaging time, $\TT \to \infty$, and the probe bandwidth, $\omega_\qp \to 0$ (in this order), the exact LDOS would be recovered.

\subsection{Many-body LDOS}

\begin{figure}
\includegraphics[width=\linewidth]{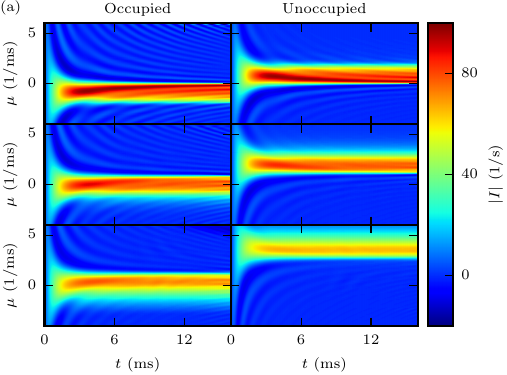}
\includegraphics[width=\linewidth]{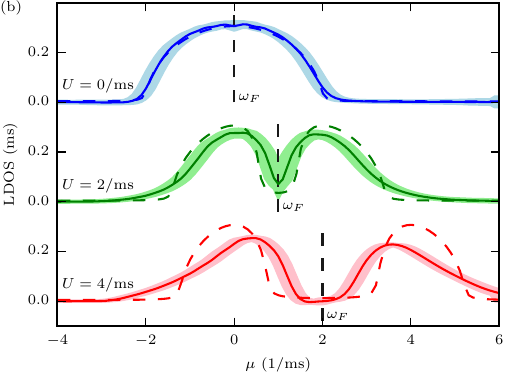}
\caption{(a) The particle current versus time for the occupied (left) and unoccupied (right) states for interaction strengths $U_i = 0$ ms$^{-1}$, $2$ ms$^{-1}$, $4$ ms$^{-1}$ (from top to bottom), and total size $N=32$, found using time-dependent density matrix renormalization group calculations (tDMRG). (b) The resulting LDOS are shown as solid lines computed through the same procedure as in Fig.~\ref{fig:avecurrnonint} and with the shaded regions indicating the standard deviation of the current combined with the probe broadening. Note that a steady-state current still forms as the system ceases to be a true insulator once it is connected to the non-interacting probe. As $U$ increases, a gap opens up between the occupied and unoccupied bands, in addition to causing a pronounced broadening. The dashed lines on the figure indicate the LDOS found using a mean-field approximation of two coupled semi-infinite systems (see~\ref{sec:numerics} and~\ref{sec:mfapprox} for details of numerical calculations). \label{fig:avecurrintmott}}
\end{figure}

We now apply the same approach to an interacting system with a constant $U_i = U$.  Figure~\ref{fig:avecurrintmott} shows the LDOS of a Mott-insulator like state, using numerical many-body calculations. As the interaction strength increases, the band splits and a gap forms between the occupied and unoccupied bands, as is typical for a Mott insulator. However, the Fermi level, $\omega_F \approx U/2$, and the occupied band is shifted to higher frequency. The observations also agree well with the approximate predictions from Green's function calculations (see Appendix~\ref{sec:mfapprox}).

Moreover, the energy-resolved local density of states elucidates the role of interactions on the physical response. For instance, a filling-dependent, conducting-to-nonconducting transition occurs as a function of $U$ for interaction-induced transport~\cite{chien2013interaction}: An inhomogeneous quench in $U$, where the interaction strength is taken from $0$ to a finite value for half the lattice, drives particles from that half of the lattice to the other (non-interacting) half so long as $U$ is not too strong and the filling not too large. Figure \ref{fig:avecurrintmott}(b) demonstrates that it is this shift of the occupied bands to higher frequency that aligns occupied states in the interacting side to open states in the non-interacting side, allowing particles to flow. As $U$ is increased further, eventually only the tail of the occupied band is aligned with open states, thus giving a decreasing---but non-zero---current solely due to many-body interactions. The mean-field solution~\cite{chien2013interaction}, however, predicts the current should go exactly to zero.  

\begin{figure}
\includegraphics[width=\linewidth]{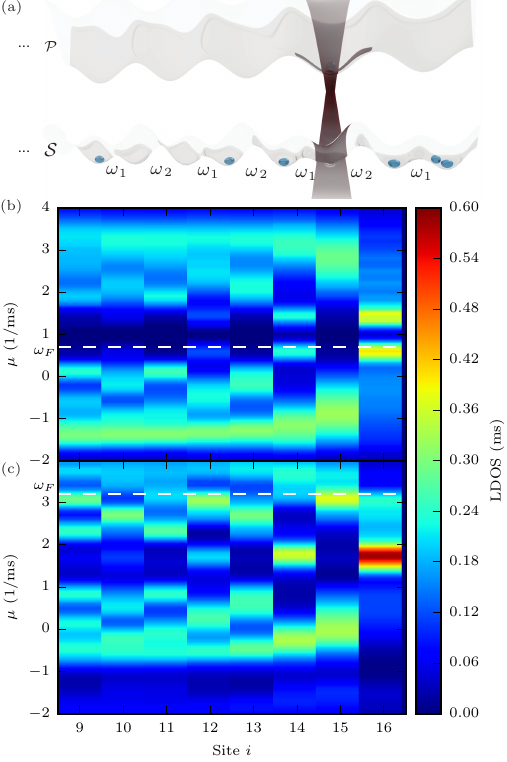}
\caption{(a) Schematic of a one-dimensional dimerized lattice system $\qs$ ---the SSH model with interactions---with alternating hopping coefficients $\omega_{1}$ and $\omega_{2}$ being interrogated by the narrow band probe $\qp$. (b,c) LDOS of the interacting SSH model with $\omega_{1} = 1$ ms$^{-1}$, $\omega_{2} = 1.5$ ms$^{-1}$, and $U = 2$~ms$^{-1}$ as a function of lattice site and chemical potential. When the lattice terminates on a weak bond, edge modes appear and live in the gap between the electronic bands. The LDOS is for (b) a half-filled system and (c) a three-quarters filled system, both of which show the presence of edge modes, found in the same manner as in Fig.~\ref{fig:avecurrintmott}. In the former case, the edge modes are half-occupied in the ground state and the interaction term is essentially equivalent to an onsite energy shift. In the latter, the larger overall filling causes a higher occupation in these same modes and broadens them into the upper band. The probe allows a clear visualization of the sublattice localization of the edge modes. The Fermi level is found empirically from where the full and empty bands overlap. \label{fig:toposurf}}
\end{figure}

When the spatial dependence of the density of states is of interest, the probe can be used in a regime analogous to a ``scanning-mode'': The probe can be coupled to the system at any lattice site, as shown in Fig.~\ref{fig:toposurf}(a). As an example, we examine the Su-Schrieffer-Heeger (SSH) model of electrons hopping in polyacetylene~\cite{SSH79,np-9-795} in the presence of many-body interactions. The SSH model has alternating electronic hopping coefficients that dimerize the lattice. Here, $\omega_{i}$ is $\omega_1 (\omega_2)$ when $i \in \qs$ is even (odd). In the non-interacting case, it has a topological invariant---the winding number. When this number is 1, edge states will be present at the boundary. When the lattice is half-filled, the presence of a homogeneous interaction $U$ does not remove the edge states. These can be seen in Figs.~\ref{fig:toposurf}(b) and (c) as the sharp, localized peak of the LDOS at the boundary, and the interaction acts by simply shifting the LDOS. When the filling is increased to three-quarters, the LDOS peak at the boundary is broadened into the upper band. Moreover, the splitting that occurs for half-filling is washed out as the Fermi frequency increases beyond the band gap.

\subsection{LDOS of an inhomogeneous interaction}

\begin{figure}
\includegraphics[width=\linewidth]{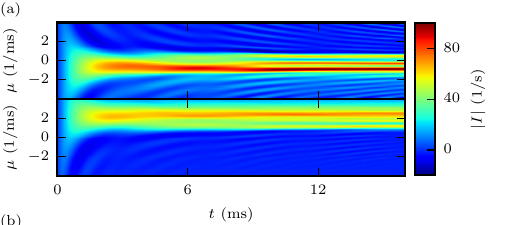}
\includegraphics[width=\linewidth]{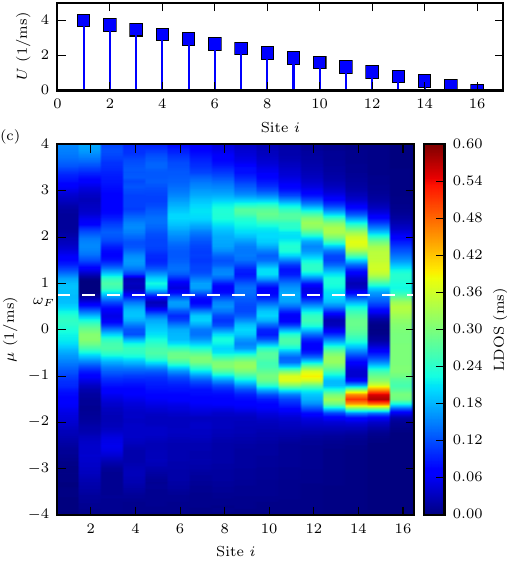}
\caption{
Scanning-mode applied to an inhomogeneous system. (a) The time-dependent current shows the formation of a steady-state flow into $\qp$ when attached to the center of $\qs$. (b) $\qs$ has an inhomogeneous interaction profile: a linear decrease from $U=4$~ms$^{-1}$ on one end to $U=0$~ms$^{-1}$ on the other. (c) The LDOS for occupied and unoccupied states as a function of frequency offset $\mu$ and contact lattice position $i$. The weakly-interacting region of the system allows more particles to occupy lower frequency states, while the more strongly-interacting side forces the open states well beyond the Fermi level. In addition, a superimposed even-odd effect is visible, which is due to the finite size of the lattice, creating oscillations from the boundary. \label{fig:avecurrintinhomo}}
\end{figure}

In previous sections, we focused on systems that have a uniform interaction term $U$ applied to the entirety of $\qs$. As an additional example, we examine a system in scanning-mode with spatially inhomogeneous interactions---e.g., a linear decrease in the interaction strength $U$, Fig.~\ref{fig:avecurrintinhomo}(b)---can determine both how the particle density shifts in space and energy. Figure~\ref{fig:avecurrintinhomo}(c) shows the occupied and unoccupied LDOS of this inhomogeneous lattice as a function of position. The spatial decrease of the interactions forces particles to the region with small interactions, where at the very end the lattice has an LDOS similar to a non-interacting system. Just near the non-interacting boundary, however, a large peak in occupied density of states forms, i.e., states pinned well below the Fermi level. On the interacting side, the number of particles is small with an LDOS just below the Fermi level. A superimposed even-odd effect is visible, which is due to finite lattice effects, creating oscillations away from the boundaries.

\begin{figure}
\includegraphics[width=\linewidth]{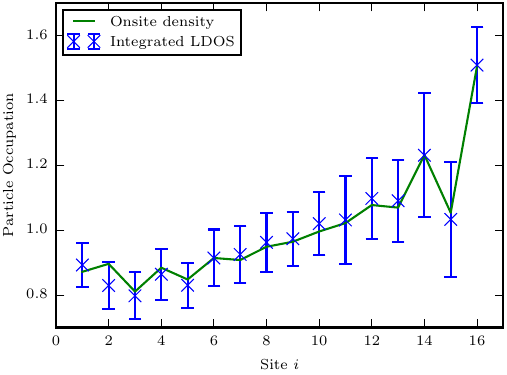}
\caption{
The integral of the measured LDOS from Fig.~\ref{fig:avecurrintinhomo} (blue x's) below the Fermi level gives the total particle occupation, which matches that found from a direct measurement of $n$, the occupation number, from the numerical calculation of the ground state (green line). The formation of a steady state current then provides not only a measure of the occupied and unoccupied states but also yields the real space occupation. The error bars represent the propagated error from the broadening and standard deviation as in the previous figures. \label{fig:occupation}}
\end{figure}

Up to the calculated error, there is a direct correspondence between the particle occupation and the integrated occupied LDOS, as with the typical non-interacting LDOS. Figure~\ref{fig:occupation} shows the comparison between the particle occupation from the simulation and the LDOS integrated over the full energy range from the application of the scanning-mode configuration to the inhomogeneous, interacting system.  We note that the atomic scanning probe approach to {\em numerically} computing the local density of states gives a straightforward alternative to other numerical methods~\cite{schoenauer2017observation}.

\section{An implementation example with ultracold strontium in optical lattices} \label{sec:realize}

Among several embodiments of the energy-resolved atomic scanning probe with ultracold atoms, such as one using nano-patterned magnetic traps, we consider as an example a highly-controllable system based on engineered dynamically-generated optical potentials with alkaline-earths, specifically, fermionic strontium atoms, $^{87}$Sr ~\cite{stellmer2014}. In the setup introduced in the above, we consider the electronic ground state $^1S_0$ and the metastable state $^3P_0$ coupled by a clock laser~\cite{s-341-632}, or a targeted three-photon process~\cite{PhysRevA.93.053417}. The system (probe) lattice is tuned to the magic-zero wavelength of the metastable (ground) state~\cite{prl-101-170504, pra-92-040501}. Therefore, atoms of $\qs$ in the ground state do not see the lattice of $\qp$, and, vice versa, atoms of $\qp$ in the metastable state do not see the lattice of $\qs$. This guarantees independent control of both the system and probe trapping potentials and interactions. Controllably overlapping the system, probe and coupling light fields can be readily implemented with feedback-controlled tweezers and lattices generated by deformable micro-mirror devices and spatial light modulators~\cite{s-354-1021, s-354-1024}. As a first demonstration, and to avoid weak interactions between atoms in the ground state, we consider implementing this probe for weakly-interacting models. Additional weak, but unwanted, interactions between atoms in the probe and the system can be reduced by a spatial offset between the probe and system sites, albeit with an increased tunneling time.

Multiple energy-resolved atomic scanning probes (tweezers) could be potentially implemented, and each probing performed multiple times, which is technically demanding but not fundamentally limited. For example, a 1D lattice system $\qs$ could be probed by multiple $\qp_i$ aligned perpendicular to the system. For fermionic strontium atoms, the number of atoms in each probe $\qp_i$ could be measured via fluorescence in a cycling transition as follows~\cite{prl-101-170504}. First, carry out a Raman transfer of the atoms in $\qp_i$ from the state 5s5p$^3P_0$ to 5s5p$^3P_2$ via 5s6s$^3S_1$. Second, implement pulsed Raman sideband imaging on the cycling transition 5s5p$^3P_2$-5s6d$^3D_3$ as in Ref.~\cite{prl-114-213002} but using the Zeeman manifold~\cite{prl-110-070403}. Third, bring atoms back to the original probe state 5s5p$^3P_0$ with a Raman transfer as in the first step. At this point, the probe $\qp_i$ could be brought again into contact with the system. 

A key requirement of the scanning probe is the ability to tune $\mu$, which for our probe is determined by the particle density as well as the lattice and external trapping potentials. Experimentally produced degenerate gases usually have a chemical potential of about 1~kHz, as considered in the calculations above, but it has been tuned from zero to 20~kHz~\cite{potnaos-113-8144}. By using engineered optical potentials, we can expand the degree of control and range of our probe by enabling different lattice shapes (per lattice site) between the probe and the system, so that the ground state levels are different between the two. The energy difference acts as an onsite energy, giving a probe offset/change in chemical potential.

Finally, we note that in this work we focus on probing the LDOS by zero-temperature transport. At finite temperatures in experiments, the Fermi-Dirac distribution will enter the expression of the particle current in Eq.~\eqref{eq:propcurr}. Therefore, extracting the LDOS from the current needs to account for this thermal broadening.

\section{Conclusion}

We conclude by noting that the density of states is a central concept in our description of matter. We have provided an operational definition of the LDOS for many-body systems, applicable in and out of equilibrium (e.g., exciting $\qs$ via a quench or some other process), to fermionic or bosonic systems, etc. The core principle is that for a current to flow---whether in a steady state or not---into an empty, narrow probe band, there must be occupied states at that energy (similarly for a full narrow band and unoccupied states). In contrast to other methods (such as the single-site method from Ref.~\onlinecite{kantian2015lattice}), this approach uses the restriction of the probe to access the long-time properties of the total current and does not require time-dependent variation once the particles are in motion. The measurement is resolved in {\it both} energy {\it and} space, as well in other characteristics (e.g., spin-resolved). We demonstrated that a cold-atom setup will allow for the measurement of this operational definition: The many-body LDOS can be extracted with minimal disturbance to the system. While cold-atom systems allow for tunability, issues still can arise regarding, e.g., the orbital character of the local states, higher energy excitations, or system-probe coupling (of which there are typically ways to treat it more accurately~\cite{Jaksch1998, Walters2013}). Unlike solid-state systems, however, the effect of these issues can be separated or even corrected in this setup. 

A related approach would be to use tunable cold-atom systems to more controllably---i.e., with less disturbance to the native state---implement Eqs.~\eqref{eq:STM1} and \eqref{eq:STM2}. The energy-resolved scanning probe, however, acts not to mimic solid-state systems, but rather to implement the ideal, tunneling-based probe, one that minimizes the total current flowing and other disturbances. This will complement the quantum-gas microscope \cite{Bakr09}, which resolves the spatial location of atoms. The flourishing of quantum simulations, from emulating condensed matter \cite{bloch_quantum_2012,cirac_goals_2012} to the physics of the early universe \cite{hung2013cosmology, kasamatsu2013atomic}, demonstrates the need to probe---both experimentally and numerically---the undisturbed density of states with spatial and energy resolution up to many-body scales. In this vein, the energy-resolved atomic scanning probe will illuminate the nature of excitations and symmetry breaking in everything from the mundane to the exotic.

\begin{acknowledgments}
Daniel Gruss acknowledges support under the Cooperative Research Agreement between the University of Maryland and the National Institute of Standards and Technology Center for Nanoscale Science and Technology, Award 70NANB14H209, through the University of Maryland. Massimiliano Di Ventra acknowledges support from the DOE Grant No. DE-FG02-05ER46204.
\end{acknowledgments}

\appendix

\renewcommand\thefigure{\Alph{section}\arabic{figure}}
\renewcommand\theequation{\Alph{section}\arabic{equation}}
\renewcommand\thesection{Appendix \Alph{section}}
\setcounter{figure}{0}

\section{Error quantification} \label{sec:numerics}

In order to extract the LDOS from real-time measurements on finite lattices, the current must be averaged over a finite time, i.e., the current will be in a quasi-steady state~\cite{di2008electrical}. The estimate of the LDOS is thus 
\begin{equation}
D(\mu)  \propto \frac{1}{\TT} \int_\qt I(t) dt .
\end{equation}
As $\TT,N \to \infty$, this will converge to the true steady state. As we demonstrate, an accurate LDOS is already apparent for small lattices and short times, and thus it requires only modest resources (it is not experimentally or numerically taxing). 

The numerical calculations are as follows: For the non-interacting system, we integrate the equations of motion to find the current~\cite{Zwolak04-1,chien2012bosonic,chien2014landauer}. The transient current when the probe ``comes in contact'' with the system is damped on the characteristic tunneling time and the recurrence time is proportional to the lattice size, which dictates both the lower and upper limits to the time region $\qt$ given a finite lattice length. We use $\qt=[ 2$~ms, $N/2$~ms$]$, where $N$ is the lattice length. We can also define an error for the non-interacting case,  
\begin{equation}
\text{Percentage Error} = 100 \; \% \cdot \frac{\sqrt{\int [D(\mu) - \dex(\mu)]^2 \; d\mu}}{\sqrt{\int \dex(\mu)^2 \; d\mu}} ,
\end{equation}
where $\dex(\mu)$ is the exact LDOS.

The uncertainty due to probe broadening is due to contributions to the current from anywhere within the probe bandwidth of $4\omega_\qp$, which results in an error, $\sigma_+(\mu)$, of
\begin{equation}
D(\mu) - \max \left[ D(\mu), D(\mu-2\omega_\qp), D(\mu+2\omega_\qp) \right]
\end{equation}
for the positive $\mu$ side and, $\sigma_-(\mu)$,
\begin{equation}
D(\mu) - \min \left[ D(\mu), D(\mu-2\omega_\qp), D(\mu+2\omega_\qp) \right]
\end{equation}
for the negative side. We combine this with the standard deviation, $\sigma_\text{stdev}$, from the time-dependent current, giving a total error of
\begin{equation} \label{eq:toterror}
\sigma^\pm_\text{tot} = \sqrt{\sigma_\pm^2 + \sigma_\text{stdev}^2} .
\end{equation}

For interacting systems, we perform time-dependent, density matrix renormalization group calculations \cite{vidal2003efficient, vidal2004efficient} within the ITensor tensor product library \cite{itensorsite}. In all simulations, we decrease the time step until the calculation converges with respect to energy and we allow the matrix product bond dimension to increase without bound. The energy cutoff is $10^{-9} \omega_\qs$. The averaging is in the region $\qt=[ 2$~ms, $N/2$~ms$]$, as with the non-interacting case. 

\section{Non-interacting LDOS} \label{sec:nonintldos}

\begin{figure}
\includegraphics[width=\linewidth]{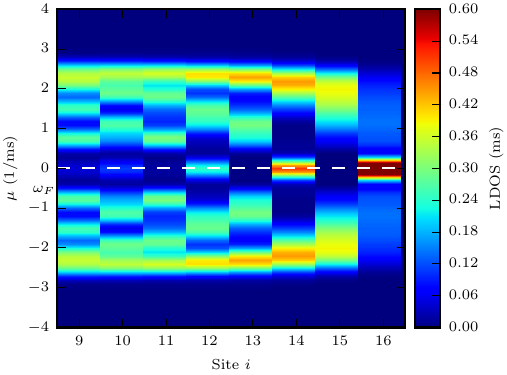}
\caption{The topological system from Fig.~\ref{fig:toposurf} with interaction strength $U=0$~ms$^{-1}$ and at half filling. In this case, the even-sites can be exactly half-occupied, removing the shift on the Fermi frequency and the splitting of the modes when half-filled. If the filling is increased or decreased, the structure of the LDOS remains the same. \label{fig:noninttopo}}
\end{figure}

The non-interacting lattice is exactly solvable for both the particle current and the LDOS. For simplicity, we take the probing lattice to be a paradigmatic, one-dimensional lattice with homogeneous hopping frequency $\omega_{\qp}$. Since we are interested in a narrow band reservoir, we take the weak hopping $\omega_{\qp}$ to also be the coupling between $\qs$ and $\qp$. For the current, we need the retarded Green's function for this semi-infinite lattice with a frequency offset $\mu$ (i.e., an onsite potential shift), which is~\cite{zwolak2002dna, chien2014landauer}
\begin{equation}\label{eq:nonintgr}
g^r_{\qp}(\omega) = \frac{1}{2 \omega_{\qp}^2} \left[ (\omega-\mu) - \im \sqrt{4 \omega_{\qp}^2 - (\omega-\mu)^2 } \right] .
\end{equation}
Using this expression, the particle current for the infinite lattice, $N \to \infty$, is given by the Landauer formula~\cite{di2008electrical},
\begin{equation}
I = -\frac{1}{2 \pi} \int_{-\infty}^{\infty} d\omega \left[ f_\qs(\omega) - f_\qp(\omega) \right] T(\omega),
\end{equation}
where $f_{\qs (\qp)}$ are the initial particle distributions (Fermi-Dirac distributions or completely filled/empty) in $\qs (\qp)$ and $T(\omega)$ is the transmission coefficient
\begin{equation}
T(\omega)=\frac{4 \Re \left[ \sqrt{4 \omega_\qp^2 - (\omega - \mu)^2} \right]
                  \Re \left[ \sqrt{4 \omega_\qs^2 - \omega^2} \right]}
    {\left| \mu + \im  \sqrt{4 \omega_\qp^2 - (\omega-\mu)^2 } + \im \sqrt{4 \omega_\qs^2 - \omega^2 } \right|^2} .
\end{equation}
Implementing the setup with $f_\qp (\omega) \in \{0, 1\}$ gives the reconstructed LDOS as simply a sum of the occupied and unoccupied states, and also directly yields the Fermi level, $\omega_F$. Above, the system of interest $\qs$ is half filled and in its zero temperature ground state. Here, we show, for simplicity, how the LDOS of a fully filled non-interacting system can be mapped out. The current in this case is  $I=-\frac{1}{2\pi}\int_{\mu-2\omega_\qp}^{\mu+2\omega_\qp}T(\omega)d\omega\approx -(2/\pi) \omega_\qp T(\mu)$ when $\omega_\qp \ll \omega_\qs$. Thus, 
\begin{equation}
I \approx -\frac{4}{\pi} \left(\frac{\omega_\qp}{\omega_\qs}\right)^2 \sqrt{4 \omega_\qs^2 - \mu^2}
\end{equation}
for $| \mu | \lesssim \omega_\qs$. Then Eq.~\eqref{eq:propcurr} gives the exact LDOS, $\dex(\mu)=\sqrt{4 \omega_\qs^2 - \mu^2}/(2\pi\omega_\qs^2)$, for a non-interacting $\qs$. The remaining terms in the full expression broaden the reconstructed LDOS by approximately the probe bandwidth, $4\omega_\qp$, which thus has to be small enough to discriminate features in the LDOS of $\qs$ and, when using a cold-atom setup, the bandwidth should be large enough to get an appreciable current. 

For systems in the thermodynamic limit, the LDOS is related to the real-space Green's function by~\cite{RickayzenBook}
\begin{equation}\label{eq:LDOSGreen}
D(\mathbf{r},\omega)=-\frac{1}{\pi} \Im \left[ g^r(\mathbf{r},\mathbf{r},\omega) \right].
\end{equation}
These two equations give the same expression for a noninteracting semi-infinite lattice. The energy dispersion is  $\omega_k=-2\omega_\qs\cos[k\pi/(N+1)]$ and $\langle r=0|\phi_k\rangle=\sqrt{2/(N+1)}\sin[k\pi/(N+1)]$ with $k=0,1,\cdots, N$. Thus, $(L=N+1)$
\begin{align}
D(r=0,\omega) &= \int dk \frac{2}{L}\sin^2\left(\frac{k\pi}{L}\right)\frac{L\delta(k-k_\qs)}{2\pi\omega_\qs\sin(\frac{k\pi}{L})}\notag \\
&=\frac{1}{2\pi\omega_\qs^2}\sqrt{4\omega_\qs^2-\omega^2}.
\end{align}
Here, $k_\qs$ is the value when $\omega-\omega_\qs=0$ is satisfied. The Green's function is similar to Eq.~\eqref{eq:nonintgr}. Explicitly, $g^r_{\qs}(\omega) = \left( \omega - \im \sqrt{4 \omega_{\qs}^2 - \omega^2 } \right) / (2 \omega_{\qs}^2)$, so $D(r=0,\omega)=-(1/\pi)\mbox{Im}(g^r)=\sqrt{4\omega_\qs^2-\omega^2}/(2\pi\omega_{\qs}^2)$. These both agree with that obtained from Eq.~\eqref{eq:propcurr} and the Landauer formula.

Figure~\ref{fig:noninttopo} shows the topological system of dimerized lattices from Fig.~\ref{fig:toposurf}(b) without an interaction term, which yields to non-perturbed state of the SSH system.

\section{Mean-field approximation} \label{sec:mfapprox}
The Green's function of the infinite, uniform Hubbard model in the mean-field approximation is \cite{RickayzenBook}
\begin{equation} \label{eq:meanfieldpoles}
g_{inf}(k,\sigma,\omega)=\frac{\omega+\mu_\qs-U(1-n)}{(\omega+\mu_\qs-\omega_k)(\omega+\mu_\qs-U)-Un\omega_k}.
\end{equation}
Here, we focus on the 1D case with $\omega_k=-2\omega_\qs\cos(k \pi / L)$ and chemical potential $\mu_\qs$ representing an onsite energy shift. $\sigma=\uparrow,\downarrow$ denotes the chosen spin species, and $n$ is the  filling factor. The poles of the Green's function lead to the energy dispersion, which has the form $(\omega-\omega^{-}_k)(\omega-\omega^{+}_{k})$ with $\omega^{\pm}_{k}$ giving the two bands:
\begin{equation}
\omega^{\pm}_{k} = \frac{U}{2} - \mu_\qs + \frac{\omega_k}{2} \pm \sqrt{\frac{U^2}{4} + \left( n - \frac{1}{2} \right) U \omega_k + \frac{\omega_k^2}{4}}.
\end{equation}
The retarded Green's function in real space is $g_{inf}^r(r,\sigma,\omega)=(1/N_l)\sum_{k} e^{ikr} g_{inf}(k,\sigma,\omega+i0^{+})$, where the summation is within the first Brillouin zone and $N_l$ is the lattice size. Since the Green's function is for a uniform, translational invariant system, we may choose $r=0$. The LDOS, however, is from the Green's function of a half-infinite lattice. An infinite lattice can be thought of as an assembly of two half-infinite lattices connected by an additional central site, so we use a derivation similar to Ref.~\cite{chien2014landauer} and obtain $g_{inf}^r (r=0,\sigma,\omega)=[\omega+\mu_\qs-Un-2\omega_\qs^2g^r (r=0,\sigma,\omega)]^{-1}$. Inverting the relation, we find $g^r (r=0,\sigma,\omega)$ for a half-infinite chain, and the LDOS can be obtained from Eq.~\eqref{eq:LDOSGreen}.

In the isolated-site limit (i.e., the ``atomic limit'') where $(\omega_\qs/U)\rightarrow 0$, the Green's function is exactly solvable and on a selected site it is $g_\qs(\sigma, \omega)=\left[\frac{1-(n/2)}{\omega+\mu_\qs}+\frac{n/2}{\omega+\mu_\qs-U} \right]$. In this limit there is no distinction between an infinite lattice and a half-infinite lattice because there is no tunneling between sites. By an analytic continuation $\omega\rightarrow \omega-i0^{+}$, we obtain $g_\qs^r$. Then, $D(\sigma,\omega)=\{[1-(n/2)]\delta(\omega+\mu_\qs)+(n/2)\delta(\omega+\mu_\qs-U)\}$. Thus, there are two peaks at $-\mu_\qs$ and $-\mu_\qs+U$. Away from the isolated-site limit, the two peaks broaden into two bands separated by $U$, and this agrees with our observation shown in Fig.~\ref{fig:avecurrintmott}.

\if\bibtexbib1

\bibliography{references}

\begin{thebibliography}{52}%
\makeatletter
\providecommand \@ifxundefined [1]{%
 \@ifx{#1\undefined}
}%
\providecommand \@ifnum [1]{%
 \ifnum #1\expandafter \@firstoftwo
 \else \expandafter \@secondoftwo
 \fi
}%
\providecommand \@ifx [1]{%
 \ifx #1\expandafter \@firstoftwo
 \else \expandafter \@secondoftwo
 \fi
}%
\providecommand \natexlab [1]{#1}%
\providecommand \enquote  [1]{``#1''}%
\providecommand \bibnamefont  [1]{#1}%
\providecommand \bibfnamefont [1]{#1}%
\providecommand \citenamefont [1]{#1}%
\providecommand \href@noop [0]{\@secondoftwo}%
\providecommand \href [0]{\begingroup \@sanitize@url \@href}%
\providecommand \@href[1]{\@@startlink{#1}\@@href}%
\providecommand \@@href[1]{\endgroup#1\@@endlink}%
\providecommand \@sanitize@url [0]{\catcode `\\12\catcode `\$12\catcode
  `\&12\catcode `\#12\catcode `\^12\catcode `\_12\catcode `\%12\relax}%
\providecommand \@@startlink[1]{}%
\providecommand \@@endlink[0]{}%
\providecommand \url  [0]{\begingroup\@sanitize@url \@url }%
\providecommand \@url [1]{\endgroup\@href {#1}{\urlprefix }}%
\providecommand \urlprefix  [0]{URL }%
\providecommand \Eprint [0]{\href }%
\providecommand \doibase [0]{http://dx.doi.org/}%
\providecommand \selectlanguage [0]{\@gobble}%
\providecommand \bibinfo  [0]{\@secondoftwo}%
\providecommand \bibfield  [0]{\@secondoftwo}%
\providecommand \translation [1]{[#1]}%
\providecommand \BibitemOpen [0]{}%
\providecommand \bibitemStop [0]{}%
\providecommand \bibitemNoStop [0]{.\EOS\space}%
\providecommand \EOS [0]{\spacefactor3000\relax}%
\providecommand \BibitemShut  [1]{\csname bibitem#1\endcsname}%
\let\auto@bib@innerbib\@empty
\bibitem [{\citenamefont {Binnig}\ \emph {et~al.}(1982)\citenamefont {Binnig},
  \citenamefont {Rohrer}, \citenamefont {Gerber},\ and\ \citenamefont
  {Weibel}}]{binnig1982surface}%
  \BibitemOpen
  \bibfield  {author} {\bibinfo {author} {\bibfnamefont {G.}~\bibnamefont
  {Binnig}}, \bibinfo {author} {\bibfnamefont {H.}~\bibnamefont {Rohrer}},
  \bibinfo {author} {\bibfnamefont {C.}~\bibnamefont {Gerber}}, \ and\ \bibinfo
  {author} {\bibfnamefont {E.}~\bibnamefont {Weibel}},\ }\href@noop {}
  {\bibfield  {journal} {\bibinfo  {journal} {Phys. Rev. Lett.}\ }\textbf
  {\bibinfo {volume} {49}},\ \bibinfo {pages} {57} (\bibinfo {year}
  {1982})}\BibitemShut {NoStop}%
\bibitem [{\citenamefont {Hansma}\ and\ \citenamefont
  {Tersoff}(1987)}]{hansma1987scanning}%
  \BibitemOpen
  \bibfield  {author} {\bibinfo {author} {\bibfnamefont {P.~K.}\ \bibnamefont
  {Hansma}}\ and\ \bibinfo {author} {\bibfnamefont {J.}~\bibnamefont
  {Tersoff}},\ }\href@noop {} {\bibfield  {journal} {\bibinfo  {journal} {J.
  Appl. Phys.}\ }\textbf {\bibinfo {volume} {61}},\ \bibinfo {pages} {R1}
  (\bibinfo {year} {1987})}\BibitemShut {NoStop}%
\bibitem [{\citenamefont {Chen}(1993)}]{chen1993introduction}%
  \BibitemOpen
  \bibfield  {author} {\bibinfo {author} {\bibfnamefont {C.~J.}\ \bibnamefont
  {Chen}},\ }\href@noop {} {\emph {\bibinfo {title} {Introduction to scanning
  tunneling microscopy}}}\ (\bibinfo  {publisher} {Oxford University Press New
  York},\ \bibinfo {year} {1993})\BibitemShut {NoStop}%
\bibitem [{\citenamefont {Stroscio}\ and\ \citenamefont
  {Kaiser}(1993)}]{stroscio1993methods}%
  \BibitemOpen
  \bibfield  {author} {\bibinfo {author} {\bibfnamefont {J.}~\bibnamefont
  {Stroscio}}\ and\ \bibinfo {author} {\bibfnamefont {W.}~\bibnamefont
  {Kaiser}},\ }\href@noop {} {\bibfield  {journal} {\bibinfo  {journal}
  {Scanning Tunneling Microscopy. Academic Press, Boston}\ } (\bibinfo {year}
  {1993})}\BibitemShut {NoStop}%
\bibitem [{\citenamefont {Wiesendanger}(1994)}]{wiesendanger1994scanning}%
  \BibitemOpen
  \bibfield  {author} {\bibinfo {author} {\bibfnamefont {R.}~\bibnamefont
  {Wiesendanger}},\ }\href@noop {} {\emph {\bibinfo {title} {Scanning probe
  microscopy and spectroscopy: {M}ethods and applications}}}\ (\bibinfo
  {publisher} {Cambridge University Press},\ \bibinfo {year}
  {1994})\BibitemShut {NoStop}%
\bibitem [{\citenamefont {Bloch}\ \emph {et~al.}(2008)\citenamefont {Bloch},
  \citenamefont {Dalibard},\ and\ \citenamefont {Zwerger}}]{bloch2008many}%
  \BibitemOpen
  \bibfield  {author} {\bibinfo {author} {\bibfnamefont {I.}~\bibnamefont
  {Bloch}}, \bibinfo {author} {\bibfnamefont {J.}~\bibnamefont {Dalibard}}, \
  and\ \bibinfo {author} {\bibfnamefont {W.}~\bibnamefont {Zwerger}},\
  }\href@noop {} {\bibfield  {journal} {\bibinfo  {journal} {Rev. Mod. Phys.}\
  }\textbf {\bibinfo {volume} {80}},\ \bibinfo {pages} {885} (\bibinfo {year}
  {2008})}\BibitemShut {NoStop}%
\bibitem [{\citenamefont {Fertig}\ \emph {et~al.}(2005)\citenamefont {Fertig},
  \citenamefont {O’hara}, \citenamefont {Huckans}, \citenamefont {Rolston},
  \citenamefont {Phillips},\ and\ \citenamefont {Porto}}]{fertig2005strongly}%
  \BibitemOpen
  \bibfield  {author} {\bibinfo {author} {\bibfnamefont {C.}~\bibnamefont
  {Fertig}}, \bibinfo {author} {\bibfnamefont {K.}~\bibnamefont {O’hara}},
  \bibinfo {author} {\bibfnamefont {J.}~\bibnamefont {Huckans}}, \bibinfo
  {author} {\bibfnamefont {S.}~\bibnamefont {Rolston}}, \bibinfo {author}
  {\bibfnamefont {W.}~\bibnamefont {Phillips}}, \ and\ \bibinfo {author}
  {\bibfnamefont {J.}~\bibnamefont {Porto}},\ }\href@noop {} {\bibfield
  {journal} {\bibinfo  {journal} {Phys. Rev. Lett.}\ }\textbf {\bibinfo
  {volume} {94}},\ \bibinfo {pages} {120403} (\bibinfo {year}
  {2005})}\BibitemShut {NoStop}%
\bibitem [{\citenamefont {Strohmaier}\ \emph {et~al.}(2007)\citenamefont
  {Strohmaier}, \citenamefont {Takasu}, \citenamefont {G{\"u}nter},
  \citenamefont {J{\"o}rdens}, \citenamefont {K{\"o}hl}, \citenamefont
  {Moritz},\ and\ \citenamefont {Esslinger}}]{strohmaier2007interaction}%
  \BibitemOpen
  \bibfield  {author} {\bibinfo {author} {\bibfnamefont {N.}~\bibnamefont
  {Strohmaier}}, \bibinfo {author} {\bibfnamefont {Y.}~\bibnamefont {Takasu}},
  \bibinfo {author} {\bibfnamefont {K.}~\bibnamefont {G{\"u}nter}}, \bibinfo
  {author} {\bibfnamefont {R.}~\bibnamefont {J{\"o}rdens}}, \bibinfo {author}
  {\bibfnamefont {M.}~\bibnamefont {K{\"o}hl}}, \bibinfo {author}
  {\bibfnamefont {H.}~\bibnamefont {Moritz}}, \ and\ \bibinfo {author}
  {\bibfnamefont {T.}~\bibnamefont {Esslinger}},\ }\href@noop {} {\bibfield
  {journal} {\bibinfo  {journal} {Phys. Rev. Lett.}\ }\textbf {\bibinfo
  {volume} {99}},\ \bibinfo {pages} {220601} (\bibinfo {year}
  {2007})}\BibitemShut {NoStop}%
\bibitem [{\citenamefont {Lamacraft}\ and\ \citenamefont
  {Moore}(2012)}]{Lamacraft12}%
  \BibitemOpen
  \bibfield  {author} {\bibinfo {author} {\bibfnamefont {A.}~\bibnamefont
  {Lamacraft}}\ and\ \bibinfo {author} {\bibfnamefont {J.}~\bibnamefont
  {Moore}},\ }in\ \href@noop {} {\emph {\bibinfo {booktitle} {Ultracold bosonic
  and fermionic gases}}},\ \bibinfo {editor} {edited by\ \bibinfo {editor}
  {\bibfnamefont {K.}~\bibnamefont {Levin}}, \bibinfo {editor} {\bibfnamefont
  {A.~L.}\ \bibnamefont {Fetter}}, \ and\ \bibinfo {editor} {\bibfnamefont
  {D.}~\bibnamefont {Stamper-Kurn}}}\ (\bibinfo  {publisher} {Elsevier},\
  \bibinfo {address} {Amsterdam, The Netherlands},\ \bibinfo {year}
  {2012})\BibitemShut {NoStop}%
\bibitem [{\citenamefont {Chien}\ \emph {et~al.}(2015)\citenamefont {Chien},
  \citenamefont {Peotta},\ and\ \citenamefont {Di~Ventra}}]{chien2015quantum}%
  \BibitemOpen
  \bibfield  {author} {\bibinfo {author} {\bibfnamefont {C.-C.}\ \bibnamefont
  {Chien}}, \bibinfo {author} {\bibfnamefont {S.}~\bibnamefont {Peotta}}, \
  and\ \bibinfo {author} {\bibfnamefont {M.}~\bibnamefont {Di~Ventra}},\
  }\href@noop {} {\bibfield  {journal} {\bibinfo  {journal} {Nat. Phys.}\
  }\textbf {\bibinfo {volume} {11}},\ \bibinfo {pages} {998} (\bibinfo {year}
  {2015})}\BibitemShut {NoStop}%
\bibitem [{\citenamefont {Ott}\ \emph {et~al.}(2004)\citenamefont {Ott},
  \citenamefont {deMirandes}, \citenamefont {Ferlaino}, \citenamefont {Roati},
  \citenamefont {Modugno},\ and\ \citenamefont
  {Inguscio}}]{ott2004collisionally}%
  \BibitemOpen
  \bibfield  {author} {\bibinfo {author} {\bibfnamefont {H.}~\bibnamefont
  {Ott}}, \bibinfo {author} {\bibfnamefont {E.}~\bibnamefont {deMirandes}},
  \bibinfo {author} {\bibfnamefont {F.}~\bibnamefont {Ferlaino}}, \bibinfo
  {author} {\bibfnamefont {G.}~\bibnamefont {Roati}}, \bibinfo {author}
  {\bibfnamefont {G.}~\bibnamefont {Modugno}}, \ and\ \bibinfo {author}
  {\bibfnamefont {M.}~\bibnamefont {Inguscio}},\ }\href@noop {} {\bibfield
  {journal} {\bibinfo  {journal} {Phys. Rev. Lett.}\ }\textbf {\bibinfo
  {volume} {92}},\ \bibinfo {pages} {160601} (\bibinfo {year}
  {2004})}\BibitemShut {NoStop}%
\bibitem [{\citenamefont {G{\"u}nter}\ \emph {et~al.}(2006)\citenamefont
  {G{\"u}nter}, \citenamefont {St{\"o}ferle}, \citenamefont {Moritz},
  \citenamefont {K{\"o}hl},\ and\ \citenamefont {Esslinger}}]{gunter2006bose}%
  \BibitemOpen
  \bibfield  {author} {\bibinfo {author} {\bibfnamefont {K.}~\bibnamefont
  {G{\"u}nter}}, \bibinfo {author} {\bibfnamefont {T.}~\bibnamefont
  {St{\"o}ferle}}, \bibinfo {author} {\bibfnamefont {H.}~\bibnamefont
  {Moritz}}, \bibinfo {author} {\bibfnamefont {M.}~\bibnamefont {K{\"o}hl}}, \
  and\ \bibinfo {author} {\bibfnamefont {T.}~\bibnamefont {Esslinger}},\
  }\href@noop {} {\bibfield  {journal} {\bibinfo  {journal} {Phys. Rev. Lett.}\
  }\textbf {\bibinfo {volume} {96}},\ \bibinfo {pages} {180402} (\bibinfo
  {year} {2006})}\BibitemShut {NoStop}%
\bibitem [{\citenamefont {Chien}\ \emph {et~al.}(2012)\citenamefont {Chien},
  \citenamefont {Zwolak},\ and\ \citenamefont {Di~Ventra}}]{chien2012bosonic}%
  \BibitemOpen
  \bibfield  {author} {\bibinfo {author} {\bibfnamefont {C.-C.}\ \bibnamefont
  {Chien}}, \bibinfo {author} {\bibfnamefont {M.}~\bibnamefont {Zwolak}}, \
  and\ \bibinfo {author} {\bibfnamefont {M.}~\bibnamefont {Di~Ventra}},\
  }\href@noop {} {\bibfield  {journal} {\bibinfo  {journal} {Phys. Rev. A}\
  }\textbf {\bibinfo {volume} {85}},\ \bibinfo {pages} {041601} (\bibinfo
  {year} {2012})}\BibitemShut {NoStop}%
\bibitem [{\citenamefont {Brantut}\ \emph {et~al.}(2012)\citenamefont
  {Brantut}, \citenamefont {Meineke}, \citenamefont {Stadler}, \citenamefont
  {Krinner},\ and\ \citenamefont {Esslinger}}]{brantut2012conduction}%
  \BibitemOpen
  \bibfield  {author} {\bibinfo {author} {\bibfnamefont {J.-P.}\ \bibnamefont
  {Brantut}}, \bibinfo {author} {\bibfnamefont {J.}~\bibnamefont {Meineke}},
  \bibinfo {author} {\bibfnamefont {D.}~\bibnamefont {Stadler}}, \bibinfo
  {author} {\bibfnamefont {S.}~\bibnamefont {Krinner}}, \ and\ \bibinfo
  {author} {\bibfnamefont {T.}~\bibnamefont {Esslinger}},\ }\href@noop {}
  {\bibfield  {journal} {\bibinfo  {journal} {Science}\ }\textbf {\bibinfo
  {volume} {337}},\ \bibinfo {pages} {1069} (\bibinfo {year}
  {2012})}\BibitemShut {NoStop}%
\bibitem [{\citenamefont {Chien}\ \emph {et~al.}(2013)\citenamefont {Chien},
  \citenamefont {Gruss}, \citenamefont {Di~Ventra},\ and\ \citenamefont
  {Zwolak}}]{chien2013interaction}%
  \BibitemOpen
  \bibfield  {author} {\bibinfo {author} {\bibfnamefont {C.-C.}\ \bibnamefont
  {Chien}}, \bibinfo {author} {\bibfnamefont {D.}~\bibnamefont {Gruss}},
  \bibinfo {author} {\bibfnamefont {M.}~\bibnamefont {Di~Ventra}}, \ and\
  \bibinfo {author} {\bibfnamefont {M.}~\bibnamefont {Zwolak}},\ }\href@noop {}
  {\bibfield  {journal} {\bibinfo  {journal} {New J. Phys.}\ }\textbf {\bibinfo
  {volume} {15}},\ \bibinfo {pages} {063026} (\bibinfo {year}
  {2013})}\BibitemShut {NoStop}%
\bibitem [{\citenamefont {Chien}\ \emph {et~al.}(2014)\citenamefont {Chien},
  \citenamefont {Di~Ventra},\ and\ \citenamefont {Zwolak}}]{chien2014landauer}%
  \BibitemOpen
  \bibfield  {author} {\bibinfo {author} {\bibfnamefont {C.-C.}\ \bibnamefont
  {Chien}}, \bibinfo {author} {\bibfnamefont {M.}~\bibnamefont {Di~Ventra}}, \
  and\ \bibinfo {author} {\bibfnamefont {M.}~\bibnamefont {Zwolak}},\
  }\href@noop {} {\bibfield  {journal} {\bibinfo  {journal} {Phys. Rev. A}\
  }\textbf {\bibinfo {volume} {90}},\ \bibinfo {pages} {023624} (\bibinfo
  {year} {2014})}\BibitemShut {NoStop}%
\bibitem [{\citenamefont {Krinner}\ \emph {et~al.}(2015)\citenamefont
  {Krinner}, \citenamefont {Stadler}, \citenamefont {Husmann}, \citenamefont
  {Brantut},\ and\ \citenamefont {Esslinger}}]{krinner2015observation}%
  \BibitemOpen
  \bibfield  {author} {\bibinfo {author} {\bibfnamefont {S.}~\bibnamefont
  {Krinner}}, \bibinfo {author} {\bibfnamefont {D.}~\bibnamefont {Stadler}},
  \bibinfo {author} {\bibfnamefont {D.}~\bibnamefont {Husmann}}, \bibinfo
  {author} {\bibfnamefont {J.-P.}\ \bibnamefont {Brantut}}, \ and\ \bibinfo
  {author} {\bibfnamefont {T.}~\bibnamefont {Esslinger}},\ }\href@noop {}
  {\bibfield  {journal} {\bibinfo  {journal} {Nature}\ }\textbf {\bibinfo
  {volume} {517}},\ \bibinfo {pages} {64} (\bibinfo {year} {2015})}\BibitemShut
  {NoStop}%
\bibitem [{\citenamefont {Hedin}\ \emph {et~al.}(1970)\citenamefont {Hedin},
  \citenamefont {Lundqvist},\ and\ \citenamefont {Lundqvist}}]{Hedin70}%
  \BibitemOpen
  \bibfield  {author} {\bibinfo {author} {\bibfnamefont {L.}~\bibnamefont
  {Hedin}}, \bibinfo {author} {\bibfnamefont {B.~I.}\ \bibnamefont
  {Lundqvist}}, \ and\ \bibinfo {author} {\bibfnamefont {S.}~\bibnamefont
  {Lundqvist}},\ }\href@noop {} {\bibfield  {journal} {\bibinfo  {journal} {J.
  Res. Nat. Bur. Stand. Sect. A}\ }\textbf {\bibinfo {volume} {74A}},\ \bibinfo
  {pages} {417} (\bibinfo {year} {1970})}\BibitemShut {NoStop}%
\bibitem [{\citenamefont {Stellmer}\ \emph {et~al.}(2014)\citenamefont
  {Stellmer}, \citenamefont {Schreck},\ and\ \citenamefont
  {Killian}}]{stellmer2014}%
  \BibitemOpen
  \bibfield  {author} {\bibinfo {author} {\bibfnamefont {S.}~\bibnamefont
  {Stellmer}}, \bibinfo {author} {\bibfnamefont {F.}~\bibnamefont {Schreck}}, \
  and\ \bibinfo {author} {\bibfnamefont {T.~C.}\ \bibnamefont {Killian}},\ }in\
  \href {\doibase 10.1142/9789814590174_0001} {\emph {\bibinfo {booktitle}
  {Annual Review of Cold Atoms and Molecules}}}\ (\bibinfo  {publisher} {World
  Scientific},\ \bibinfo {year} {2014})\ pp.\ \bibinfo {pages}
  {1--80}\BibitemShut {NoStop}%
\bibitem [{\citenamefont {Daley}\ \emph {et~al.}(2008)\citenamefont {Daley},
  \citenamefont {Boyd}, \citenamefont {Ye},\ and\ \citenamefont
  {Zoller}}]{prl-101-170504}%
  \BibitemOpen
  \bibfield  {author} {\bibinfo {author} {\bibfnamefont {A.~J.}\ \bibnamefont
  {Daley}}, \bibinfo {author} {\bibfnamefont {M.~M.}\ \bibnamefont {Boyd}},
  \bibinfo {author} {\bibfnamefont {J.}~\bibnamefont {Ye}}, \ and\ \bibinfo
  {author} {\bibfnamefont {P.}~\bibnamefont {Zoller}},\ }\href@noop {}
  {\bibfield  {journal} {\bibinfo  {journal} {Phys. Rev. Lett.}\ }\textbf
  {\bibinfo {volume} {101}},\ \bibinfo {pages} {170504} (\bibinfo {year}
  {2008})}\BibitemShut {NoStop}%
\bibitem [{\citenamefont {Regal}\ and\ \citenamefont
  {Jin}(2003)}]{prl-90-230404}%
  \BibitemOpen
  \bibfield  {author} {\bibinfo {author} {\bibfnamefont {C.~A.}\ \bibnamefont
  {Regal}}\ and\ \bibinfo {author} {\bibfnamefont {D.~S.}\ \bibnamefont
  {Jin}},\ }\href {\doibase 10.1103/PhysRevLett.90.230404} {\bibfield
  {journal} {\bibinfo  {journal} {Phys. Rev. Lett.}\ }\textbf {\bibinfo
  {volume} {90}},\ \bibinfo {pages} {230404} (\bibinfo {year}
  {2003})}\BibitemShut {NoStop}%
\bibitem [{\citenamefont {Shin}\ \emph {et~al.}(2007)\citenamefont {Shin},
  \citenamefont {Schunck}, \citenamefont {Schirotzek},\ and\ \citenamefont
  {Ketterle}}]{prl-99-090403}%
  \BibitemOpen
  \bibfield  {author} {\bibinfo {author} {\bibfnamefont {Y.}~\bibnamefont
  {Shin}}, \bibinfo {author} {\bibfnamefont {C.~H.}\ \bibnamefont {Schunck}},
  \bibinfo {author} {\bibfnamefont {A.}~\bibnamefont {Schirotzek}}, \ and\
  \bibinfo {author} {\bibfnamefont {W.}~\bibnamefont {Ketterle}},\ }\href
  {\doibase 10.1103/physrevlett.99.090403} {\bibfield  {journal} {\bibinfo
  {journal} {Phys. Rev. Lett.}\ }\textbf {\bibinfo {volume} {99}},\ \bibinfo
  {pages} {090403} (\bibinfo {year} {2007})}\BibitemShut {NoStop}%
\bibitem [{\citenamefont {Barredo}\ \emph {et~al.}(2016)\citenamefont
  {Barredo}, \citenamefont {de~Léséleuc}, \citenamefont {Lienhard},
  \citenamefont {Lahaye},\ and\ \citenamefont {Browaeys}}]{s-354-1021}%
  \BibitemOpen
  \bibfield  {author} {\bibinfo {author} {\bibfnamefont {D.}~\bibnamefont
  {Barredo}}, \bibinfo {author} {\bibfnamefont {S.}~\bibnamefont
  {de~Léséleuc}}, \bibinfo {author} {\bibfnamefont {V.}~\bibnamefont
  {Lienhard}}, \bibinfo {author} {\bibfnamefont {T.}~\bibnamefont {Lahaye}}, \
  and\ \bibinfo {author} {\bibfnamefont {A.}~\bibnamefont {Browaeys}},\ }\href
  {\doibase 10.1126/science.aah3778} {\bibfield  {journal} {\bibinfo  {journal}
  {Science}\ }\textbf {\bibinfo {volume} {354}},\ \bibinfo {pages}
  {1021–1023} (\bibinfo {year} {2016})}\BibitemShut {NoStop}%
\bibitem [{\citenamefont {Endres}\ \emph {et~al.}(2016)\citenamefont {Endres},
  \citenamefont {Bernien}, \citenamefont {Keesling}, \citenamefont {Levine},
  \citenamefont {Anschuetz}, \citenamefont {Krajenbrink}, \citenamefont
  {Senko}, \citenamefont {Vuletic}, \citenamefont {Greiner},\ and\
  \citenamefont {Lukin}}]{s-354-1024}%
  \BibitemOpen
  \bibfield  {author} {\bibinfo {author} {\bibfnamefont {M.}~\bibnamefont
  {Endres}}, \bibinfo {author} {\bibfnamefont {H.}~\bibnamefont {Bernien}},
  \bibinfo {author} {\bibfnamefont {A.}~\bibnamefont {Keesling}}, \bibinfo
  {author} {\bibfnamefont {H.}~\bibnamefont {Levine}}, \bibinfo {author}
  {\bibfnamefont {E.~R.}\ \bibnamefont {Anschuetz}}, \bibinfo {author}
  {\bibfnamefont {A.}~\bibnamefont {Krajenbrink}}, \bibinfo {author}
  {\bibfnamefont {C.}~\bibnamefont {Senko}}, \bibinfo {author} {\bibfnamefont
  {V.}~\bibnamefont {Vuletic}}, \bibinfo {author} {\bibfnamefont
  {M.}~\bibnamefont {Greiner}}, \ and\ \bibinfo {author} {\bibfnamefont
  {M.~D.}\ \bibnamefont {Lukin}},\ }\href {\doibase 10.1126/science.aah3752}
  {\bibfield  {journal} {\bibinfo  {journal} {Science}\ }\textbf {\bibinfo
  {volume} {354}},\ \bibinfo {pages} {1024–1027} (\bibinfo {year}
  {2016})}\BibitemShut {NoStop}%
\bibitem [{\citenamefont {Marti}\ \emph {et~al.}(2018)\citenamefont {Marti},
  \citenamefont {Hutson}, \citenamefont {Goban}, \citenamefont {Campbell},
  \citenamefont {Poli},\ and\ \citenamefont {Ye}}]{PhysRevLett.120.103201}%
  \BibitemOpen
  \bibfield  {author} {\bibinfo {author} {\bibfnamefont {G.~E.}\ \bibnamefont
  {Marti}}, \bibinfo {author} {\bibfnamefont {R.~B.}\ \bibnamefont {Hutson}},
  \bibinfo {author} {\bibfnamefont {A.}~\bibnamefont {Goban}}, \bibinfo
  {author} {\bibfnamefont {S.~L.}\ \bibnamefont {Campbell}}, \bibinfo {author}
  {\bibfnamefont {N.}~\bibnamefont {Poli}}, \ and\ \bibinfo {author}
  {\bibfnamefont {J.}~\bibnamefont {Ye}},\ }\href {\doibase
  10.1103/PhysRevLett.120.103201} {\bibfield  {journal} {\bibinfo  {journal}
  {Phys. Rev. Lett.}\ }\textbf {\bibinfo {volume} {120}},\ \bibinfo {pages}
  {103201} (\bibinfo {year} {2018})}\BibitemShut {NoStop}%
\bibitem [{\citenamefont {Zhang}\ \emph {et~al.}(2015)\citenamefont {Zhang},
  \citenamefont {Cheng}, \citenamefont {Zhai},\ and\ \citenamefont
  {Zhang}}]{prl-115-135301}%
  \BibitemOpen
  \bibfield  {author} {\bibinfo {author} {\bibfnamefont {R.}~\bibnamefont
  {Zhang}}, \bibinfo {author} {\bibfnamefont {Y.}~\bibnamefont {Cheng}},
  \bibinfo {author} {\bibfnamefont {H.}~\bibnamefont {Zhai}}, \ and\ \bibinfo
  {author} {\bibfnamefont {P.}~\bibnamefont {Zhang}},\ }\href {\doibase
  10.1103/PhysRevLett.115.135301} {\bibfield  {journal} {\bibinfo  {journal}
  {Phys. Rev. Lett.}\ }\textbf {\bibinfo {volume} {115}},\ \bibinfo {pages}
  {135301} (\bibinfo {year} {2015})}\BibitemShut {NoStop}%
\bibitem [{\citenamefont {H{\"o}fer}\ \emph {et~al.}(2015)\citenamefont
  {H{\"o}fer}, \citenamefont {Riegger}, \citenamefont {Scazza}, \citenamefont
  {Hofrichter}, \citenamefont {Fernandes}, \citenamefont {Parish},
  \citenamefont {Levinsen}, \citenamefont {Bloch},\ and\ \citenamefont
  {F{\"o}lling}}]{prl-115-265302}%
  \BibitemOpen
  \bibfield  {author} {\bibinfo {author} {\bibfnamefont {M.}~\bibnamefont
  {H{\"o}fer}}, \bibinfo {author} {\bibfnamefont {L.}~\bibnamefont {Riegger}},
  \bibinfo {author} {\bibfnamefont {F.}~\bibnamefont {Scazza}}, \bibinfo
  {author} {\bibfnamefont {C.}~\bibnamefont {Hofrichter}}, \bibinfo {author}
  {\bibfnamefont {D.~R.}\ \bibnamefont {Fernandes}}, \bibinfo {author}
  {\bibfnamefont {M.~M.}\ \bibnamefont {Parish}}, \bibinfo {author}
  {\bibfnamefont {J.}~\bibnamefont {Levinsen}}, \bibinfo {author}
  {\bibfnamefont {I.}~\bibnamefont {Bloch}}, \ and\ \bibinfo {author}
  {\bibfnamefont {S.}~\bibnamefont {F{\"o}lling}},\ }\href {\doibase
  10.1103/PhysRevLett.115.265302} {\bibfield  {journal} {\bibinfo  {journal}
  {Phys. Rev. Lett.}\ }\textbf {\bibinfo {volume} {115}},\ \bibinfo {pages}
  {265302} (\bibinfo {year} {2015})}\BibitemShut {NoStop}%
\bibitem [{\citenamefont {Pagano}\ \emph {et~al.}(2015)\citenamefont {Pagano},
  \citenamefont {Mancini}, \citenamefont {Cappellini}, \citenamefont {Livi},
  \citenamefont {Sias}, \citenamefont {Catani}, \citenamefont {Inguscio},\ and\
  \citenamefont {Fallani}}]{prl-115-265301}%
  \BibitemOpen
  \bibfield  {author} {\bibinfo {author} {\bibfnamefont {G.}~\bibnamefont
  {Pagano}}, \bibinfo {author} {\bibfnamefont {M.}~\bibnamefont {Mancini}},
  \bibinfo {author} {\bibfnamefont {G.}~\bibnamefont {Cappellini}}, \bibinfo
  {author} {\bibfnamefont {L.}~\bibnamefont {Livi}}, \bibinfo {author}
  {\bibfnamefont {C.}~\bibnamefont {Sias}}, \bibinfo {author} {\bibfnamefont
  {J.}~\bibnamefont {Catani}}, \bibinfo {author} {\bibfnamefont
  {M.}~\bibnamefont {Inguscio}}, \ and\ \bibinfo {author} {\bibfnamefont
  {L.}~\bibnamefont {Fallani}},\ }\href {\doibase
  10.1103/PhysRevLett.115.265301} {\bibfield  {journal} {\bibinfo  {journal}
  {Phys. Rev. Lett.}\ }\textbf {\bibinfo {volume} {115}},\ \bibinfo {pages}
  {265301} (\bibinfo {year} {2015})}\BibitemShut {NoStop}%
\bibitem [{\citenamefont {Zwolak}\ and\ \citenamefont
  {Vidal}(2004)}]{Zwolak04-1}%
  \BibitemOpen
  \bibfield  {author} {\bibinfo {author} {\bibfnamefont {M.}~\bibnamefont
  {Zwolak}}\ and\ \bibinfo {author} {\bibfnamefont {G.}~\bibnamefont {Vidal}},\
  }\href@noop {} {\bibfield  {journal} {\bibinfo  {journal} {Phys. Rev. Lett.}\
  }\textbf {\bibinfo {volume} {93}},\ \bibinfo {pages} {207205} (\bibinfo
  {year} {2004})}\BibitemShut {NoStop}%
\bibitem [{\citenamefont {Su}\ \emph {et~al.}(1979)\citenamefont {Su},
  \citenamefont {Schrieffer},\ and\ \citenamefont {Heeger}}]{SSH79}%
  \BibitemOpen
  \bibfield  {author} {\bibinfo {author} {\bibfnamefont {W.~P.}\ \bibnamefont
  {Su}}, \bibinfo {author} {\bibfnamefont {J.~R.}\ \bibnamefont {Schrieffer}},
  \ and\ \bibinfo {author} {\bibfnamefont {A.~J.}\ \bibnamefont {Heeger}},\
  }\href@noop {} {\bibfield  {journal} {\bibinfo  {journal} {Phys. Rev. Lett.}\
  }\textbf {\bibinfo {volume} {42}},\ \bibinfo {pages} {1698} (\bibinfo {year}
  {1979})}\BibitemShut {NoStop}%
\bibitem [{\citenamefont {Atala}\ \emph {et~al.}(2013)\citenamefont {Atala},
  \citenamefont {Aidelsburger}, \citenamefont {Barreiro}, \citenamefont
  {Abanin}, \citenamefont {Kitagawa}, \citenamefont {Demler},\ and\
  \citenamefont {Bloch}}]{np-9-795}%
  \BibitemOpen
  \bibfield  {author} {\bibinfo {author} {\bibfnamefont {M.}~\bibnamefont
  {Atala}}, \bibinfo {author} {\bibfnamefont {M.}~\bibnamefont {Aidelsburger}},
  \bibinfo {author} {\bibfnamefont {J.~T.}\ \bibnamefont {Barreiro}}, \bibinfo
  {author} {\bibfnamefont {D.}~\bibnamefont {Abanin}}, \bibinfo {author}
  {\bibfnamefont {T.}~\bibnamefont {Kitagawa}}, \bibinfo {author}
  {\bibfnamefont {E.}~\bibnamefont {Demler}}, \ and\ \bibinfo {author}
  {\bibfnamefont {I.}~\bibnamefont {Bloch}},\ }\href {\doibase
  10.1038/nphys2790} {\bibfield  {journal} {\bibinfo  {journal} {Nature Phys.}\
  }\textbf {\bibinfo {volume} {9}},\ \bibinfo {pages} {795} (\bibinfo {year}
  {2013})}\BibitemShut {NoStop}%
\bibitem [{\citenamefont {Schoenauer}\ \emph {et~al.}(2017)\citenamefont
  {Schoenauer}, \citenamefont {Schmitteckert},\ and\ \citenamefont
  {Schuricht}}]{schoenauer2017observation}%
  \BibitemOpen
  \bibfield  {author} {\bibinfo {author} {\bibfnamefont {B.}~\bibnamefont
  {Schoenauer}}, \bibinfo {author} {\bibfnamefont {P.}~\bibnamefont
  {Schmitteckert}}, \ and\ \bibinfo {author} {\bibfnamefont {D.}~\bibnamefont
  {Schuricht}},\ }\href@noop {} {\bibfield  {journal} {\bibinfo  {journal}
  {Phys. Rev. B}\ }\textbf {\bibinfo {volume} {95}},\ \bibinfo {pages} {205103}
  (\bibinfo {year} {2017})}\BibitemShut {NoStop}%
\bibitem [{\citenamefont {Martin}\ \emph {et~al.}(2013)\citenamefont {Martin},
  \citenamefont {Bishof}, \citenamefont {Swallows}, \citenamefont {Zhang},
  \citenamefont {Benko}, \citenamefont {von Stecher}, \citenamefont {Gorshkov},
  \citenamefont {Rey},\ and\ \citenamefont {Ye}}]{s-341-632}%
  \BibitemOpen
  \bibfield  {author} {\bibinfo {author} {\bibfnamefont {M.~J.}\ \bibnamefont
  {Martin}}, \bibinfo {author} {\bibfnamefont {M.}~\bibnamefont {Bishof}},
  \bibinfo {author} {\bibfnamefont {M.~D.}\ \bibnamefont {Swallows}}, \bibinfo
  {author} {\bibfnamefont {X.}~\bibnamefont {Zhang}}, \bibinfo {author}
  {\bibfnamefont {C.}~\bibnamefont {Benko}}, \bibinfo {author} {\bibfnamefont
  {J.}~\bibnamefont {von Stecher}}, \bibinfo {author} {\bibfnamefont {A.~V.}\
  \bibnamefont {Gorshkov}}, \bibinfo {author} {\bibfnamefont {A.~M.}\
  \bibnamefont {Rey}}, \ and\ \bibinfo {author} {\bibfnamefont
  {J.}~\bibnamefont {Ye}},\ }\href {\doibase 10.1126/science.1236929}
  {\bibfield  {journal} {\bibinfo  {journal} {Science}\ }\textbf {\bibinfo
  {volume} {341}},\ \bibinfo {pages} {632} (\bibinfo {year}
  {2013})}\BibitemShut {NoStop}%
\bibitem [{\citenamefont {Barker}\ \emph {et~al.}(2016)\citenamefont {Barker},
  \citenamefont {Pisenti}, \citenamefont {Reschovsky},\ and\ \citenamefont
  {Campbell}}]{PhysRevA.93.053417}%
  \BibitemOpen
  \bibfield  {author} {\bibinfo {author} {\bibfnamefont {D.~S.}\ \bibnamefont
  {Barker}}, \bibinfo {author} {\bibfnamefont {N.~C.}\ \bibnamefont {Pisenti}},
  \bibinfo {author} {\bibfnamefont {B.~J.}\ \bibnamefont {Reschovsky}}, \ and\
  \bibinfo {author} {\bibfnamefont {G.~K.}\ \bibnamefont {Campbell}},\ }\href
  {\doibase 10.1103/PhysRevA.93.053417} {\bibfield  {journal} {\bibinfo
  {journal} {Phys. Rev. A}\ }\textbf {\bibinfo {volume} {93}},\ \bibinfo
  {pages} {053417} (\bibinfo {year} {2016})}\BibitemShut {NoStop}%
\bibitem [{\citenamefont {Safronova}\ \emph {et~al.}(2015)\citenamefont
  {Safronova}, \citenamefont {Zuhrianda}, \citenamefont {Safronova},\ and\
  \citenamefont {Clark}}]{pra-92-040501}%
  \BibitemOpen
  \bibfield  {author} {\bibinfo {author} {\bibfnamefont {M.~S.}\ \bibnamefont
  {Safronova}}, \bibinfo {author} {\bibfnamefont {Z.}~\bibnamefont
  {Zuhrianda}}, \bibinfo {author} {\bibfnamefont {U.~I.}\ \bibnamefont
  {Safronova}}, \ and\ \bibinfo {author} {\bibfnamefont {C.~W.}\ \bibnamefont
  {Clark}},\ }\href {\doibase 10.1103/physreva.92.040501} {\bibfield  {journal}
  {\bibinfo  {journal} {Phys. Rev. A}\ }\textbf {\bibinfo {volume} {92}},\
  \bibinfo {pages} {040501} (\bibinfo {year} {2015})}\BibitemShut {NoStop}%
\bibitem [{\citenamefont {Parsons}\ \emph {et~al.}(2015)\citenamefont
  {Parsons}, \citenamefont {Huber}, \citenamefont {Mazurenko}, \citenamefont
  {Chiu}, \citenamefont {Setiawan}, \citenamefont {Wooley-Brown}, \citenamefont
  {Blatt},\ and\ \citenamefont {Greiner}}]{prl-114-213002}%
  \BibitemOpen
  \bibfield  {author} {\bibinfo {author} {\bibfnamefont {M.~F.}\ \bibnamefont
  {Parsons}}, \bibinfo {author} {\bibfnamefont {F.}~\bibnamefont {Huber}},
  \bibinfo {author} {\bibfnamefont {A.}~\bibnamefont {Mazurenko}}, \bibinfo
  {author} {\bibfnamefont {C.~S.}\ \bibnamefont {Chiu}}, \bibinfo {author}
  {\bibfnamefont {W.}~\bibnamefont {Setiawan}}, \bibinfo {author}
  {\bibfnamefont {K.}~\bibnamefont {Wooley-Brown}}, \bibinfo {author}
  {\bibfnamefont {S.}~\bibnamefont {Blatt}}, \ and\ \bibinfo {author}
  {\bibfnamefont {M.}~\bibnamefont {Greiner}},\ }\href {\doibase
  10.1103/physrevlett.114.213002} {\bibfield  {journal} {\bibinfo  {journal}
  {Phys. Rev. Lett.}\ }\textbf {\bibinfo {volume} {114}},\ \bibinfo {pages}
  {213002} (\bibinfo {year} {2015})}\BibitemShut {NoStop}%
\bibitem [{\citenamefont {Schindler}\ \emph {et~al.}(2013)\citenamefont
  {Schindler}, \citenamefont {Monz}, \citenamefont {Nigg}, \citenamefont
  {Barreiro}, \citenamefont {Martinez}, \citenamefont {Brandl}, \citenamefont
  {Chwalla}, \citenamefont {Hennrich},\ and\ \citenamefont
  {Blatt}}]{prl-110-070403}%
  \BibitemOpen
  \bibfield  {author} {\bibinfo {author} {\bibfnamefont {P.}~\bibnamefont
  {Schindler}}, \bibinfo {author} {\bibfnamefont {T.}~\bibnamefont {Monz}},
  \bibinfo {author} {\bibfnamefont {D.}~\bibnamefont {Nigg}}, \bibinfo {author}
  {\bibfnamefont {J.~T.}\ \bibnamefont {Barreiro}}, \bibinfo {author}
  {\bibfnamefont {E.~A.}\ \bibnamefont {Martinez}}, \bibinfo {author}
  {\bibfnamefont {M.~F.}\ \bibnamefont {Brandl}}, \bibinfo {author}
  {\bibfnamefont {M.}~\bibnamefont {Chwalla}}, \bibinfo {author} {\bibfnamefont
  {M.}~\bibnamefont {Hennrich}}, \ and\ \bibinfo {author} {\bibfnamefont
  {R.}~\bibnamefont {Blatt}},\ }\href {\doibase 10.1103/PhysRevLett.110.070403}
  {\bibfield  {journal} {\bibinfo  {journal} {Phys. Rev. Lett.}\ }\textbf
  {\bibinfo {volume} {110}},\ \bibinfo {pages} {070403} (\bibinfo {year}
  {2013})}\BibitemShut {NoStop}%
\bibitem [{\citenamefont {Krinner}\ \emph {et~al.}(2016)\citenamefont
  {Krinner}, \citenamefont {Lebrat}, \citenamefont {Husmann}, \citenamefont
  {Grenier}, \citenamefont {Brantut},\ and\ \citenamefont
  {Esslinger}}]{potnaos-113-8144}%
  \BibitemOpen
  \bibfield  {author} {\bibinfo {author} {\bibfnamefont {S.}~\bibnamefont
  {Krinner}}, \bibinfo {author} {\bibfnamefont {M.}~\bibnamefont {Lebrat}},
  \bibinfo {author} {\bibfnamefont {D.}~\bibnamefont {Husmann}}, \bibinfo
  {author} {\bibfnamefont {C.}~\bibnamefont {Grenier}}, \bibinfo {author}
  {\bibfnamefont {J.-P.}\ \bibnamefont {Brantut}}, \ and\ \bibinfo {author}
  {\bibfnamefont {T.}~\bibnamefont {Esslinger}},\ }\href {\doibase
  10.1073/pnas.1601812113} {\bibfield  {journal} {\bibinfo  {journal}
  {Proceedings of the National Academy of Sciences}\ }\textbf {\bibinfo
  {volume} {113}},\ \bibinfo {pages} {8144} (\bibinfo {year}
  {2016})}\BibitemShut {NoStop}%
\bibitem [{\citenamefont {Kantian}\ \emph {et~al.}(2015)\citenamefont
  {Kantian}, \citenamefont {Schollw{\"o}ck},\ and\ \citenamefont
  {Giamarchi}}]{kantian2015lattice}%
  \BibitemOpen
  \bibfield  {author} {\bibinfo {author} {\bibfnamefont {A.}~\bibnamefont
  {Kantian}}, \bibinfo {author} {\bibfnamefont {U.}~\bibnamefont
  {Schollw{\"o}ck}}, \ and\ \bibinfo {author} {\bibfnamefont {T.}~\bibnamefont
  {Giamarchi}},\ }\href@noop {} {\bibfield  {journal} {\bibinfo  {journal}
  {Phys. Rev. Lett.}\ }\textbf {\bibinfo {volume} {115}},\ \bibinfo {pages}
  {165301} (\bibinfo {year} {2015})}\BibitemShut {NoStop}%
\bibitem [{\citenamefont {Jaksch}\ \emph {et~al.}(1998)\citenamefont {Jaksch},
  \citenamefont {Bruder}, \citenamefont {Cirac}, \citenamefont {Gardiner},\
  and\ \citenamefont {Zoller}}]{Jaksch1998}%
  \BibitemOpen
  \bibfield  {author} {\bibinfo {author} {\bibfnamefont {D.}~\bibnamefont
  {Jaksch}}, \bibinfo {author} {\bibfnamefont {C.}~\bibnamefont {Bruder}},
  \bibinfo {author} {\bibfnamefont {J.~I.}\ \bibnamefont {Cirac}}, \bibinfo
  {author} {\bibfnamefont {C.~W.}\ \bibnamefont {Gardiner}}, \ and\ \bibinfo
  {author} {\bibfnamefont {P.}~\bibnamefont {Zoller}},\ }\href@noop {}
  {\bibfield  {journal} {\bibinfo  {journal} {Phys. Rev. Lett.}\ }\textbf
  {\bibinfo {volume} {81}},\ \bibinfo {pages} {3108} (\bibinfo {year}
  {1998})}\BibitemShut {NoStop}%
\bibitem [{\citenamefont {Walters}\ \emph {et~al.}(2013)\citenamefont
  {Walters}, \citenamefont {Cotugno}, \citenamefont {Johnson}, \citenamefont
  {Clark},\ and\ \citenamefont {Jaksch}}]{Walters2013}%
  \BibitemOpen
  \bibfield  {author} {\bibinfo {author} {\bibfnamefont {R.}~\bibnamefont
  {Walters}}, \bibinfo {author} {\bibfnamefont {G.}~\bibnamefont {Cotugno}},
  \bibinfo {author} {\bibfnamefont {T.~H.}\ \bibnamefont {Johnson}}, \bibinfo
  {author} {\bibfnamefont {S.~R.}\ \bibnamefont {Clark}}, \ and\ \bibinfo
  {author} {\bibfnamefont {D.}~\bibnamefont {Jaksch}},\ }\href@noop {}
  {\bibfield  {journal} {\bibinfo  {journal} {Phys. Rev. A}\ }\textbf {\bibinfo
  {volume} {87}},\ \bibinfo {pages} {043613} (\bibinfo {year}
  {2013})}\BibitemShut {NoStop}%
\bibitem [{\citenamefont {Bakr}\ \emph {et~al.}(2009)\citenamefont {Bakr},
  \citenamefont {Gillen}, \citenamefont {Peng}, \citenamefont {Folling},\ and\
  \citenamefont {Greiner}}]{Bakr09}%
  \BibitemOpen
  \bibfield  {author} {\bibinfo {author} {\bibfnamefont {W.~S.}\ \bibnamefont
  {Bakr}}, \bibinfo {author} {\bibfnamefont {J.~I.}\ \bibnamefont {Gillen}},
  \bibinfo {author} {\bibfnamefont {A.}~\bibnamefont {Peng}}, \bibinfo {author}
  {\bibfnamefont {S.}~\bibnamefont {Folling}}, \ and\ \bibinfo {author}
  {\bibfnamefont {M.}~\bibnamefont {Greiner}},\ }\href@noop {} {\bibfield
  {journal} {\bibinfo  {journal} {Nature}\ }\textbf {\bibinfo {volume} {462}},\
  \bibinfo {pages} {74} (\bibinfo {year} {2009})}\BibitemShut {NoStop}%
\bibitem [{\citenamefont {Bloch}\ \emph {et~al.}(2012)\citenamefont {Bloch},
  \citenamefont {Dalibard},\ and\ \citenamefont
  {Nascimbène}}]{bloch_quantum_2012}%
  \BibitemOpen
  \bibfield  {author} {\bibinfo {author} {\bibfnamefont {I.}~\bibnamefont
  {Bloch}}, \bibinfo {author} {\bibfnamefont {J.}~\bibnamefont {Dalibard}}, \
  and\ \bibinfo {author} {\bibfnamefont {S.}~\bibnamefont {Nascimbène}},\
  }\href@noop {} {\bibfield  {journal} {\bibinfo  {journal} {Nat. Phys.}\
  }\textbf {\bibinfo {volume} {8}},\ \bibinfo {pages} {267} (\bibinfo {year}
  {2012})}\BibitemShut {NoStop}%
\bibitem [{\citenamefont {Cirac}\ and\ \citenamefont
  {Zoller}(2012)}]{cirac_goals_2012}%
  \BibitemOpen
  \bibfield  {author} {\bibinfo {author} {\bibfnamefont {J.~I.}\ \bibnamefont
  {Cirac}}\ and\ \bibinfo {author} {\bibfnamefont {P.}~\bibnamefont {Zoller}},\
  }\href@noop {} {\bibfield  {journal} {\bibinfo  {journal} {Nat. Phys.}\
  }\textbf {\bibinfo {volume} {8}},\ \bibinfo {pages} {264} (\bibinfo {year}
  {2012})}\BibitemShut {NoStop}%
\bibitem [{\citenamefont {Hung}\ \emph {et~al.}(2013)\citenamefont {Hung},
  \citenamefont {Gurarie},\ and\ \citenamefont {Chin}}]{hung2013cosmology}%
  \BibitemOpen
  \bibfield  {author} {\bibinfo {author} {\bibfnamefont {C.-L.}\ \bibnamefont
  {Hung}}, \bibinfo {author} {\bibfnamefont {V.}~\bibnamefont {Gurarie}}, \
  and\ \bibinfo {author} {\bibfnamefont {C.}~\bibnamefont {Chin}},\ }\href@noop
  {} {\bibfield  {journal} {\bibinfo  {journal} {Science}\ }\textbf {\bibinfo
  {volume} {341}},\ \bibinfo {pages} {1213} (\bibinfo {year}
  {2013})}\BibitemShut {NoStop}%
\bibitem [{\citenamefont {Kasamatsu}\ \emph {et~al.}(2013)\citenamefont
  {Kasamatsu}, \citenamefont {Ichinose},\ and\ \citenamefont
  {Matsui}}]{kasamatsu2013atomic}%
  \BibitemOpen
  \bibfield  {author} {\bibinfo {author} {\bibfnamefont {K.}~\bibnamefont
  {Kasamatsu}}, \bibinfo {author} {\bibfnamefont {I.}~\bibnamefont {Ichinose}},
  \ and\ \bibinfo {author} {\bibfnamefont {T.}~\bibnamefont {Matsui}},\
  }\href@noop {} {\bibfield  {journal} {\bibinfo  {journal} {Phys. Rev. Lett.}\
  }\textbf {\bibinfo {volume} {111}},\ \bibinfo {pages} {115303} (\bibinfo
  {year} {2013})}\BibitemShut {NoStop}%
\bibitem [{\citenamefont {Di~Ventra}(2008)}]{di2008electrical}%
  \BibitemOpen
  \bibfield  {author} {\bibinfo {author} {\bibfnamefont {M.}~\bibnamefont
  {Di~Ventra}},\ }\href@noop {} {\emph {\bibinfo {title} {Electrical transport
  in nanoscale systems}}}\ (\bibinfo  {publisher} {Cambridge University Press
  Cambridge},\ \bibinfo {address} {Cambridge, UK},\ \bibinfo {year}
  {2008})\BibitemShut {NoStop}%
\bibitem [{\citenamefont {Vidal}(2003)}]{vidal2003efficient}%
  \BibitemOpen
  \bibfield  {author} {\bibinfo {author} {\bibfnamefont {G.}~\bibnamefont
  {Vidal}},\ }\href@noop {} {\bibfield  {journal} {\bibinfo  {journal} {Phys.
  Rev. Lett.}\ }\textbf {\bibinfo {volume} {91}},\ \bibinfo {pages} {147902}
  (\bibinfo {year} {2003})}\BibitemShut {NoStop}%
\bibitem [{\citenamefont {Vidal}(2004)}]{vidal2004efficient}%
  \BibitemOpen
  \bibfield  {author} {\bibinfo {author} {\bibfnamefont {G.}~\bibnamefont
  {Vidal}},\ }\href@noop {} {\bibfield  {journal} {\bibinfo  {journal} {Phys.
  Rev. Lett.}\ }\textbf {\bibinfo {volume} {93}},\ \bibinfo {pages} {040502}
  (\bibinfo {year} {2004})}\BibitemShut {NoStop}%
\bibitem [{\citenamefont {Stoudenmire}\ and\ \citenamefont
  {White}()}]{itensorsite}%
  \BibitemOpen
  \bibfield  {author} {\bibinfo {author} {\bibfnamefont {E.~M.}\ \bibnamefont
  {Stoudenmire}}\ and\ \bibinfo {author} {\bibfnamefont {S.~R.}\ \bibnamefont
  {White}},\ }\href@noop {} {\enquote {\bibinfo {title} {{ITensor} -
  {Intelligent} {Tensor} {Library}},}\ }\bibinfo {howpublished}
  {\url{http://itensor.org/}},\ \bibinfo {note} {accessed:
  2016-04-06}\BibitemShut {NoStop}%
\bibitem [{\citenamefont {Zwolak}\ and\ \citenamefont
  {Di~Ventra}(2002)}]{zwolak2002dna}%
  \BibitemOpen
  \bibfield  {author} {\bibinfo {author} {\bibfnamefont {M.}~\bibnamefont
  {Zwolak}}\ and\ \bibinfo {author} {\bibfnamefont {M.}~\bibnamefont
  {Di~Ventra}},\ }\href@noop {} {\bibfield  {journal} {\bibinfo  {journal}
  {Appl. Phys. Lett.}\ }\textbf {\bibinfo {volume} {81}},\ \bibinfo {pages}
  {925} (\bibinfo {year} {2002})}\BibitemShut {NoStop}%
\bibitem [{\citenamefont {Rickayzen}(2013)}]{RickayzenBook}%
  \BibitemOpen
  \bibfield  {author} {\bibinfo {author} {\bibfnamefont {G.}~\bibnamefont
  {Rickayzen}},\ }\href@noop {} {\emph {\bibinfo {title} {Green's functions and
  condensed matter}}}\ (\bibinfo  {publisher} {Dover Publications, Inc.,
  Mineola, New York},\ \bibinfo {year} {2013})\BibitemShut {NoStop}%
\end{thebibliography}

\else

\fi

\end{document}